\documentclass[12pt]{article}
\usepackage{jheppub}
\usepackage{mathtools}
\usepackage{enumerate}
\usepackage{bm}
\usepackage{dsfont}
\usepackage{scalerel}

\def\para{\paragraph{}}
\def\be#1\ee{\begin{align}#1\end{align}}
\def\nn{\nonumber}
\def\eqn{\eqref}

\newcommand{\Tr}{\mathrm{Tr}}
\newcommand{\vol}{\mathrm{Vol}}
\newcommand{\detprime}{\mathrm{det} {}^\prime}
\newcommand{\bra}[1]{\langle{#1}|}
\newcommand{\ket}[1]{|{#1}\rangle}
\newcommand{\braket}[2]{\langle{#1}|{#2}\rangle}

\newcommand{\kket}[1]{\Vert{#1}\rangle\!\rangle}
\newcommand{\classv}{\mathcal{V}}
\newcommand{\classa}{\mathcal{A}}
\newcommand{\Icurr}{\mathcal{J}}
\newcommand{\LambdaR}{\Lambda[\mathcal{R}]}
\newcommand{\LambdaRdash}{\Lambda[\mathcal{R}']}
\newcommand{\LambdaRRdash}{\Lambda[\mathcal{R}, \mathcal{R}']}

\DeclareSymbolFont{myletters}{OML}{ztmcm}{m}{it}
\DeclareMathSymbol{\uplambda}{\mathord}{myletters}{"15}

\title{Boundary States for Chiral Symmetries in Two Dimensions}

\author{Philip Boyle Smith}
\author{and David Tong}

\affiliation{
Department of Applied Mathematics and Theoretical Physics, \\
University of Cambridge, Cambridge, CB3 OWA, UK
}

\emailAdd{pb594@damtp.cam.ac.uk}
\emailAdd{d.tong@damtp.cam.ac.uk}

\abstract{We study boundary states for Dirac fermions in $d=1+1$ dimensions that preserve Abelian chiral symmetries, meaning that the left- and right-moving fermions carry different charges. We derive simple expressions, in terms of the fermion charge assignments, for the boundary central charge and for the ground state degeneracy of the system when two different boundary conditions are imposed at either end of an interval. We show that all such boundary states fall into one of two classes, related to SPT phases supported by $(-1)^F$, which are characterised by the existence of an unpaired Majorana zero mode.}

\begin{document}

\setcounter{tocdepth}{2}
\maketitle

\newpage
\section{Introduction}

When can a quantum field theory be placed on a manifold with boundary? And what symmetries must be sacrificed in the process? Questions of this kind have played a prominent role in the developing story of topological phases of matter \cite{senthil}.

\para
A partial answer to these questions is provided by the observation that symmetries with a 't Hooft anomaly do not fare well in the presence of a boundary. Specifically, a theory with a 't Hooft anomaly for some symmetry $G$ cannot be placed on a manifold with boundary while preserving $G$.

\para
Some intuition for this statement comes from SPT phases. A $(d+1)$-dimensional SPT phase, protected by some symmetry $G$, has the property that, when placed on a manifold with boundary, its $d$-dimensional edge modes exhibit a 't Hooft anomaly for $G$. Conversely, any theory with a 't Hooft anomaly for $G$ can be realised as the boundary of a higher dimensional SPT phase. The simple observation that $\partial^2=0$ means that the boundary theory cannot, itself, be placed on a manifold with boundary \cite{ryu2}. Indeed, the authors of \cite{jensen} proved in a large number of cases that a symmetry $G$ that suffers a 't Hooft anomaly cannot be preserved in the presence of a boundary. (See also \cite{luckock} for earlier work.)

\para
Our interest in this paper lies in the possibility of preserving chiral symmetries in the presence of a boundary. These symmetries do not suffer from 't Hooft anomalies, but the anomaly cancels in an interesting way which means that it's not entirely obvious how to impose boundary conditions that are consistent with the symmetry. A particularly interesting example of this phenomenon is provided by the Standard Model: is it possible to place the Standard Model on a manifold with boundary without explicitly breaking the chiral electroweak symmetry? To our knowledge, it is not presently known how to do this.

\para
Here we take baby steps. We explore the boundary conditions for Dirac fermions in $d=1+1$ dimensions, where we have the language of boundary conformal field theory at our disposal. We construct the most general boundary state consistent with specified chiral, Abelian symmetries and determine a number of properties of these states. We will explain our main results later in this introduction, but first it will prove useful to give a simple example to set the scene.

\subsection{A Simple Example: A Single Fermion}

We can illustrate some of these issues by looking at a single Dirac fermion in $d=1+1$ dimensions.
A single Dirac fermion exhibits a $U(1)_V \times U(1)_A$ symmetry. Neither the vector nor axial symmetry has a 't Hooft anomaly, but there is a mixed anomaly between them. This suggests that we should be able to impose boundary conditions that preserve either $U(1)_V$ or $U(1)_A$, but not both.

\para
Indeed, it is not difficult to write down classes of boundary conditions that relate the left-moving fermion $\psi_L$ to the right-moving fermion $\psi_R$ and do the job. We could, for example, consider the boundary condition
\be V[\theta]: \ \ \psi_L = e^{i\theta} \psi_R \label{v}\ee
This preserves the vector symmetry $U(1)_V$ at the expense of the axial symmetry $U(1)_A$. The boundary condition depends on a phase $e^{i\theta}$, whose existence can be traced to the broken $U(1)_A$.

\para
Alternatively, we could impose the boundary condition
\be A[\theta]: \ \ \psi_L = e^{i\theta} \psi_R^\dagger \label{a}\ee
This now preserves the axial symmetry but breaks the vector. In the context of condensed matter physics, this axial boundary condition is referred to as {\it Andreev reflection}. Physically, an electron bounces off the boundary and returns as a hole, a phenomenon that is seen when a wire is attached to a superconductor. Again, the boundary condition is parametrised by a phase.

\subsection*{Compatibility of Boundary Conditions}

Our primary interest in this paper is in theories that live on an interval. If we attempt to impose different boundary conditions on each end, there are a number of questions that arise. Most importantly, we can ask: is the resulting theory consistent? If it is, we can also ask: how many ground states does the theory have?

\para
The essential physics can already be seen in the single Dirac fermion. At each end, we get a choice of vector \eqn{v} or axial \eqn{a} boundary condition, each specified by a phase, $\theta_1$ at one end and $\theta_2$ at the other. There are two possibilities for the resulting physics:
\begin{itemize}
\item $V[\theta_1]-V[\theta_2]$ or $A[\theta_1]-A[\theta_2]$: With $VV$ or $AA$ boundary conditions, the system generically has a single ground state. However, in the special case that $\theta_1=\theta_2$, the Dirac fermion has a single complex zero mode. This increases the ground state degeneracy to 2.
\item $A[\theta_1]-V[\theta_2]$ or $V[\theta_1]-A[\theta_2]$: With mixed $AV$ or $V\hspace{-.25em}A$ boundary conditions, there is a single Majorana zero mode\footnote{To see this, it is simplest to split each Weyl fermion into its Majorana-Weyl components: $\psi_L = \chi_L^1 + i\chi_L^2$ and $\psi_R = \chi_R^1 + i\chi_R^2$. A constant spinor is compatible with the two boundary conditions \eqn{v} and \eqn{a} only if
\be
\left(\begin{array}{c}\chi^1_L \\ \chi^2_L\end{array}\right)
=
R[-\theta_1]
\left(\begin{array}{cc}1 & 0 \\ 0 & -1\end{array}\right)
R[\theta_2]
\left(\begin{array}{c}\chi^1_L \\ \chi^2_L\end{array}\right)
\nn\ee
where $R[\theta]$ is the $2\times 2$ matrix that implements a rotation by $\theta$. But the combination of these three matrices is a reflection about some axis and so always has a real eigenvector with eigenvalue $+1$. The same argument applied to the $VV$ and $AA$ case gives a rotation matrix $R[\theta_2-\theta_1]$ which has eigenvalue $+1$ only when $\theta_1=\theta_2$.} for all $\theta_1$ and $\theta_2$.

A single, quantum mechanical Majorana mode is a particularly simple example of an anomalous quantum system. Perhaps the quickest way to see this is to note that a single Majorana zero mode contributes $\sqrt{2}$ to the counting of states in the partition function. We learn that while both $V$ and $A$ boundary conditions are possible, they are not mutually compatible.

\end{itemize}

\subsection{Summary of Results}

The story described above becomes more complicated when we have two or more fermions. This is because there are now non-anomalous chiral symmetries where it is less obvious how to implement the boundary condition. 

\para
For example, consider two free Dirac fermions. We may wish to place the theory on a manifold with boundary, now preserving the $U(1)$ global symmetry under which the two left-moving fermions have charges $+3$ and $+4$, and the two right-moving fermions have charges $+5$ and $0$. This symmetry does not suffer a 't Hooft anomaly, by virtue of the fact that
\be 3^2 + 4^2 = 5^2 + 0^2 \label{345}\ee
Yet any linear boundary condition, like \eqn{v} or \eqn{a}, relating left- and right-moving fermions will not respect this symmetry.

\para
In such situations, there are a number of ways to proceed. One could incorporate additional degrees of freedom on the boundary such that it is possible to write down boundary conditions that are linear in the fermions but continue to respect the symmetry. The fermion-rotor model of \cite{joe} provides an example of this kind.

\para
Alternatively, one could attempt to quantise the theory by imposing the non-linear boundary condition $J^\mu n_\mu=0$ where $J^\mu$ is the current and $n^\mu$ is normal to the boundary. As far as we're aware, it is not known how to do this in higher dimensions. However, in $d=1+1$, the formalism of boundary conformal field theory allows one to proceed in this manner. The purpose of this paper is to understand some of the properties of boundaries that preserve chiral symmetries like \eqn{345}.

\para
Specifically, we will consider $N$ Dirac fermions and, on a given boundary, insist that a $U(1)^N$ subgroup of the chiral symmetry is preserved. Here we would like to advertise our two main results. For this, we first need to introduce a little notation.

\para
We assign the left-moving fermions charges $Q_{\alpha, i}$ and the right-moving fermions charges $\bar{Q}_{\alpha i}$, where $\alpha=1,\ldots,N$ labels the $U(1)$ symmetry, and $i=1,\ldots,N$ labels the fermion. Typically, these charges differ so that we are dealing with a chiral symmetry. We insist that these symmetries do not suffer from mixed 't Hooft anomalies, which means that our charge matrices must obey the constraints
\be Q_{\alpha i} Q_{\beta i} = \bar{Q}_{\alpha i} \bar{Q}_{\beta i} \label{earlyanom}\ee
From these charge matrices, we can build a rational orthogonal matrix
\be \mathcal{R}_{ij} = (\bar{Q}^{-1})_{i\alpha} Q_{\alpha j} \nn\ee
The choice of such a matrix specifies the $U(1)^N$ symmetry that is preserved by the boundary. A general boundary state is then characterised by a choice of $\mathcal{R}$, together with a bunch of phases that are analogous to the $e^{i\theta}$ factors that we met in \eqn{v} and \eqn{a}. 

\para
One final piece of notation: to each charge matrix $\mathcal{R}$ we can associate a lattice $\Lambda[\mathcal{R}] \subseteq \mathbb{Z}^N$. This lattice consists of all integer-valued vectors, $\lambda_i\in \mathbb{Z}$ which satisfy
\be \LambdaR = \Big\{\, \lambda \in \mathbb{Z}^N \ : \ \mathcal{R} \lambda \in \mathbb{Z}^N \ \Big\} \nn\ee
Now we are in a position to describe our results. The first is a simple expression for the Affleck-Ludwig boundary central charge \cite{afflud}; we show that this is given by
\be g_\mathcal{R} = \sqrt{\vol(\LambdaR)}\label{introg}\ee
where $\vol(\LambdaR)$ is the volume of the primitive unit cell of the lattice $\Lambda$. The same result, in a rather different context, can be found in \cite{bachas}.

\para
If each fermion is given a simple boundary condition \eqn{v} or \eqn{a}, it is simple to check that $g_\mathcal{R} = 1$. More complicated, chiral boundary conditions have $g_\mathcal{R}>1$. Typically $g_\mathcal{R}$ is not an integer.

\para
Our second result is concerned with the situation in which we place the fermions on an interval, with different symmetries $\mathcal{R}$ and $\mathcal{R}'$ preserved at the two ends. In this case, we derive an elegant formula for the number of ground states $G[\mathcal{R},\mathcal{R}'] $ of the system. For generic values of the phases, we find
\be G[\mathcal{R},\mathcal{R}'] = \frac{\sqrt{\vol(\LambdaR) \, \vol(\LambdaRdash)}}{\vol(\LambdaRRdash)} \sqrt{\detprime(\mathds{1} - \mathcal{R}^T \mathcal{R}')} \label{introG}\ee
where the intersection lattice $\LambdaRRdash$ is defined to be those integer vectors $\lambda$ which obey $\mathcal{R} \lambda = \mathcal{R}' \lambda \in \mathbb{Z}^N$. For special values of the phases, the ground state degeneracy can be enhanced in way that we detail in the text.

\para
It is not at all obvious that the expression for ground state degeneracy $G[\mathcal{R},\mathcal{R}']$ is an integer. In fact, we claim that $G[\mathcal{R},\mathcal{R}']$ is either an integer, or is $\sqrt{2}$ times an integer,
\be G[\mathcal{R},\mathcal{R}'] \in \mathbb{Z} \cup \sqrt{2} \, \mathbb{Z} \label{intro2}\ee
The case of $\sqrt{2} \, \mathbb{Z}$ is telling us that the system has an unpaired Majorana zero mode, and hence the two boundary conditions are mutually incompatible. Indeed, related factors of $\sqrt{2}$ have appeared in the early study of non-BPS D-branes \cite{sen,witten} and, more recently, in the analysis of SPT phases \cite{dw,yuji}.

\para
Furthermore, we show that all symmetries $\mathcal{R}$ fall into one of two classes which, following the discussion of a single fermion above, we denote as class $\classv$ and class $\classa$. Any choice of boundary conditions $\mathcal{R}$ and $\mathcal{R}'$ from within the same class result in an integer ground state degeneracy. In contrast, if $\mathcal{R}$ and $\mathcal{R}'$ are chosen from different classes, then there is an unpaired Majorana zero mode.

\subsection*{The Relationship to Gapped Systems}

As stressed in \cite{ryu2}, there is a close correspondence between ways to put a theory on a manifold with boundary, and ways to gap a theory preserving certain symmetries. The intuitive correspondence is that, given any interaction that gaps the system, one can turn it on in the Lagrangian with a spatial, step-function profile. At low energies, then this then looks like a boundary condition for the massless fields. In the context of the Standard Model, the question becomes: is it possible to gap the fermions \emph{without} breaking electroweak symmetry? Perturbatively, the answer to this question is famously ``no". Non-perturbatively, things are far less clear.

\para
For the $d=1+1$ chiral symmetries considered in this paper, there is a long literature devoted to the question of when these systems can be gapped, starting with the influential work of Haldane \cite{haldane}. (See, for example, \cite{saulina,wangwen,levin} for further developments.) It was shown in \cite{ryu2} that the possible boundary states that one can build are entirely equivalent to Haldane's so-called ``null vector condition"\footnote{Since we are dealing with Dirac fermions, viewed as edge modes they have a trivial K-matrix, $K={\rm diag}(\mathds{1}_N,-\mathds{1}_N)$. Applied to this case, Haldane's criterion simply states that it's possible to find a gapping potential (albeit one which is typically irrelevant) provided that the charge vectors $l_{\alpha i} = (Q_{\alpha i},-\bar{Q}_{\alpha i})$ obey $l_{\alpha i}K_{ij} l_{\beta j}=0$, which is simply the anomaly condition \eqn{earlyanom}.}.

\para
When the boundary condition is viewed as a gapped phase, the two classes $\classv$ and $\classa$ that we described above translate into a $\mathbb{Z}_2$ classification of SPT phases, protected by $(-1)^F$. The question of whether there is an SPT interpretation of the full ground state degeneracy \eqn{introG} remains open.

\subsubsection*{The Plan of the Paper}

In Section \ref{bssec}, we give a review of the boundary conformal field theory techniques that we use, and construct the boundary states preserving a given $U(1)^N$ symmetry. We give a partial proof that the boundary central charge is given by \eqn{introg}. This proof is completed in Section \ref{gsec} where we consider theories on an interval, with different boundary conditions imposed at each end. 

\para
We also derive the formula \eqn{introG} in Section \ref{gsec}. Most of the effort is taken up with the showing that, for large classes of examples, the ground state degeneracy obeys \eqn{intro2}, with all states falling into one of two classes. (This is far from trivial and there remain a number of special cases where we have been unable to prove the result, but have compelling numerical evidence.)

\para
Finally, in Section \ref{examplesec}, we give a number of examples of boundary conditions. We also include several appendices which detail technical results that are omitted from the main text.

\section{Construction of Boundary States} \label{bssec}

In this section we construct all possible boundary conditions that one can impose on $N$ Dirac fermions in $d=1+1$ dimensions, subject to the requirement that there is vanishing flux of both energy and of a chosen $U(1)^N$ current flowing into the boundary. The boundary conformal field theory techniques we use are standard, and consist of first finding Ishibashi states, then imposing both clustering and the Cardy condition.

\subsection{Boundary Conformal Field Theory}

Our setting is a two-dimensional conformal field theory. We would like to place this system on an interval. In doing so, we must impose boundary conditions $A$ and $B$ at either end. We would like to understand what our options are for these boundary conditions. Furthermore, for fixed boundary conditions, we would like to understand the content of the Hilbert space $\mathcal{H}_{AB}$ of the resulting theory. The answers to both these questions can be found in the framework of boundary conformal field theory, first introduced by Cardy \cite{john}. Reviews of this topic can be found, for example, in \cite{cardy,matthias}.

\para
The key idea is to use modular covariance or, what string theorists refer to as open/closed string duality. The content of the Hilbert space $\mathcal{H}_{AB}$ is encoded in the partition function $\Tr_{\mathcal{H}_{AB}}(e^{-\beta H_{AB}})$, evaluated with antiperiodic boundary conditions on Euclidean time $\beta$. Here, both the Hilbert space $\mathcal{H}_{AB}$ and the Hamiltonian $H_{AB}$ depend on the conditions imposed on each boundary.

\para
Open-closed duality then states that the partition function $\Tr_{\mathcal{H}_{AB}}(e^{-\beta H_{AB}})$ on the interval can be related to the Hamiltonian $H_P$ of the system defined on an anti-periodic circle,
\be {\Tr_{\mathcal{H}_{AB}}(e^{-\beta H_{AB}})} \quad = {\bra{B} e^{-L H_P} \ket{A}} \label{openclosed}\ee
Or, pictorially,
\be \vcenter{\hbox{\includegraphics[scale=.8]{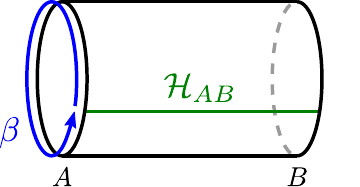}}} = \vcenter{\hbox{\includegraphics[scale=.8]{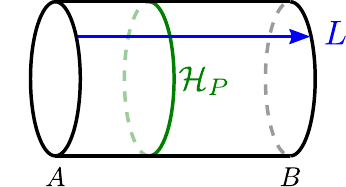}}} \nn\ee
This relates the open string partition function to a matrix element between two states. The states $\ket{A}$ and $\ket{B}$ live in the closed string Hilbert space $\mathcal{H}_P$ of the system on an anti-periodic circle, and are known as boundary states, or \emph{Cardy} states: they must obey a number of properties that we describe below.

\para
To make use of the machinery of 2D CFT, we rewrite both sides of \eqn{openclosed} by conformally mapping them into a planar geometry. The left hand side of \eqn{openclosed} equals the partition function on a half-annulus, while the right equals that of a full annulus:
\be {\Tr_{\mathcal{H}_{AB}}((e^{-\pi \beta / L})^{L_0 - \frac{c}{24}})} = {\bra{B} (e^{-4\pi L/\beta})^{\frac{1}{2}(L_0 + \bar{L}_0 - \frac{c}{12})} \ket{A}} \label{annuli}\ee
Or, in pictures,
\be \vcenter{\hbox{\includegraphics[scale=.8]{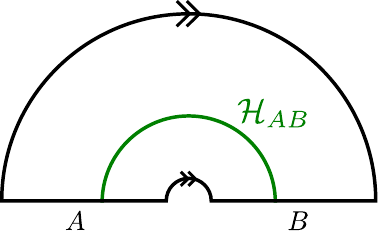}}} \ \ = \ \ \vcenter{\hbox{\includegraphics[scale=.8]{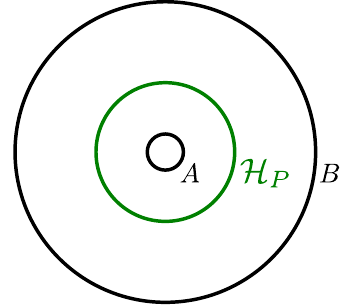}}} \nn\ee
Here, the $L_n$ on the left are the single set of Virasoro generators appropriate to a half-plane, while on the right there are both $L_n$ and $\bar{L}_n$. A crucial point is that in transforming the right hand side, the antiperiodic circle in \eqn{openclosed} maps to a \emph{periodic} annulus in \eqn{annuli}, thus finally earning the name $\mathcal{H}_P$ for this Hilbert space.

\para
All we shall need from \eqn{annuli} can be expressed in a succinct, easy-to-use form as follows. First we define the partition functions
\be
\text{open-sector:} \qquad \mathcal{Z}_{AB}(q) &= \Tr_{\mathcal{H}_{AB}}(q^{L_0 - \frac{c}{24}}) \label{open}\\
\text{closed-sector:} \qquad \mathcal{Z}_P(q) &= \bra{B} q^{\frac{1}{2}(L_0 + \bar{L}_0 - \frac{c}{12})} \ket{A} \label{closed}
\ee
The arguments of the partition functions in the equality \eqn{annuli} are not the same; rather, they are related by a modular transformation. In general we define the standard modular parameters $q\equiv e^{2\pi i \tau}$ and $\tilde{q}\equiv e^{2\pi i \tilde{\tau}}$, where the modular $\mathcal{S}$-transformation relates $\tilde{\tau} = -1/\tau$. We will denote this transformation by $\mathcal{S}(q) = \tilde{q}$. The equality \eqn{annuli} can then be written as
\be \mathcal{Z}_{AB}(q) = \mathcal{Z}_P(\mathcal{S}(q)) \label{cardy}\ee
This is \emph{Cardy's condition}. It allows the content of the `mystery' Hilbert space $\mathcal{H}_{AB}$ to be read off from the right-hand side, which involves a matrix element between two states in the known Hilbert space $\mathcal{H}_P$, namely that of the periodic or NS sector in the plane. Equally, the requirement that \eqn{cardy} defines a sensible partition function $\mathcal{Z}_{AB}(q)$ places strong constraints on the allowed boundary states one can impose.

\subsection{Ishibashi States for Free Fermions} \label{ishibashisec}

We now specialise to our system of interest, $N$ Dirac fermions in $d=2$ dimensions. Our convention for the action and currents can be found in Appendix \ref{fermionsappendix}. In the absence of a boundary, these fermions enjoy a $SO(2N)_L\times SO(2N)_R$ chiral symmetry. Our aim is to study boundaries that preserve some choice of subgroup
\be U(1)^N \subset SO(2N)_L \times SO(2N)_R \nn\ee
Each $U(1)_\alpha$, with $\alpha =1,\ldots,N$, is specified by the charges $Q_{\alpha i}$ for each of the $i=1,\ldots, N$ left-moving fermions and, independently, charges $\bar{Q}_{\alpha i}$ for each of the $i=1,\ldots, N$ right-moving fermions.

\para
We begin by working in the closed sector, with Hilbert space $\mathcal{H}_P$. The $\mathfrak{u}(1)^{N}$ current algebra consists of holomorphic and anti-holomorphic currents $J_i$ and $\bar{J}_i$, with $i=1,\ldots, N$, whose mode expansion is
\be [J_{i,n}, J_{j,m}] = [\bar{J}_{i,n}, \bar{J}_{j,m}] = n \delta_{ij} \delta_{n+m,0} \nn\ee
The preserved $U(1)_\alpha$ symmetries have currents
\be
\Icurr_{\alpha,n} = Q_{\alpha i} J_{i,n}
\ \ \ \text{and} \ \ \
\bar{\Icurr}_{\alpha,n} = \bar{Q}_{\alpha i} \bar{J}_{i,n} \label{curlyj}
\ee
The requirement that no $U(1)_\alpha$ current flows into the boundary amounts to saying that
\be (\Icurr_{\alpha,n} + \bar{\Icurr}_{\alpha,-n}) \, \ket{A} = 0 \label{noflow}\ee
For solutions to exist, we must have the vanishing commutator
\be [\Icurr_{\alpha,n} + \bar{\Icurr}_{\alpha,-n}, \Icurr_{\beta,m} + \bar{\Icurr}_{\beta,-m}] = n \delta_{n+m,0} (Q_{\alpha i} Q_{\beta i} - \bar{Q}_{\alpha i} \bar{Q}_{\beta i}) \nn\ee
This tells us that the charges of the left- and right-movers must satisfy the $N^2$ constraints
\be Q_{\alpha i} Q_{\beta i} = \bar{Q}_{\alpha i} \bar{Q}_{\beta i} \label{anomfree}\ee
This is precisely the requirement that there is no mixed 't Hooft anomaly between the $U(1)_\alpha$ and $U(1)_\beta$ symmetries. From now on, we assume that all such anomalies vanish.

\para
Our description of a $U(1)^N$ subgroup in terms of charges may be intuitive, but suffers from an inherent redundancy: any redefinition of the charges by
\be Q_{\alpha i} \rightarrow U_{\alpha \beta} Q_{\beta i} \qquad \bar{Q}_{\alpha i} \rightarrow U_{\alpha \beta} \bar{Q}_{\beta i} \nn\ee
with $U_{\alpha \beta}$ unimodular does not change the $U(1)^N$ subgroup they describe. One way of eliminating this redundancy is to introduce the matrix
\be \mathcal{R}_{ij} = (\bar{Q}^{-1})_{i\alpha} Q_{\alpha j} \label{R}\ee
which is rational and orthogonal. The possible anomaly-free $U(1)^N$ subgroups of $U(1)^N_L \times U(1)^N_R \subset SO(2N)_L \times SO(2N)_R$ are then in one-to-one correspondence with such matrices. For these reasons, we will use both $(Q, \bar{Q})$ and $\mathcal{R}$ in what follows when specifying the $U(1)^N$ symmetry.

\para
The construction of the boundary states requires further knowledge about the structure of $\mathcal{H}_P$. Under the current algebra generated by $J_{i,n}$ and $\bar{J}_{i,n}$, the Hilbert space decomposes into charge sectors. In each sector, there is a ground state $\ket{\lambda, \bar{\lambda}}$ with charges
\be
J_{i,0} \ket{\lambda, \bar{\lambda}} = \lambda_i \ket{\lambda, \bar{\lambda}}
\ \ \ , \ \ \
\bar{J}_{i,0} \ket{\lambda, \bar{\lambda}} = \bar{\lambda}_i \ket{\lambda, \bar{\lambda}} \label{lambda}
\ee
where $\lambda_i, \bar{\lambda}_i \in {\mathbb Z}$.\footnote{Our phase convention for the $\ket{\lambda, \bar{\lambda}}$ is detailed in Appendix~\ref{fermionsappendix}. However, in almost all of what follows, this choice will play no role.} These ground states obey $J_{i,n} \ket{\lambda, \bar{\lambda}} = \bar{J}_{i,n} \ket{\lambda, \bar{\lambda}} = 0$ for $n \geq 1$. Excitations above the ground state are then constructed by acting with $J_{i,-n}$ and $\bar{J}_{i,-n}$ for $n\geq 1$. The condition \eqn{noflow} that $U(1)_\alpha$ is preserved can be imposed as separate condition on each charge sector $(\lambda, \bar{\lambda})$, and reads
\be (\mathcal{R}_{ij} J_{j,n} + \bar{J}_{i,-n}) \, \ket{A} = 0 \label{noflow2}\ee
Importantly, not all charge sectors $(\lambda, \bar{\lambda})$ admit solutions to \eqn{noflow2}. The $n=0$ equation tells us that we must restrict to those charge sectors that obey
\be \bar{\lambda}_i = -\mathcal{R}_{ij} \lambda_j \label{qlambda}\ee
Not all $\lambda$ will give rise to integer-valued solutions of this equation. Instead, $\lambda$ must lie in a certain sub-lattice of $\mathbb{Z}^N$, defined by
\be \LambdaR = \Big\{\, \lambda \in \mathbb{Z}^N \ : \ \mathcal{R} \lambda \in \mathbb{Z}^N \ \Big\} \nn\ee
The allowed charge sectors are then $(\lambda, \bar{\lambda}) = (\lambda, -\mathcal{R} \lambda)$ for $\lambda \in \LambdaR$. In each such sector, the condition \eqn{noflow2} is solved by Ishibashi states which take the form \cite{ishibashi}
\be \kket{\lambda, \bar{\lambda}; \mathcal{R}} = \exp \left( -\sum_{n=1}^\infty \frac{1}{n} \mathcal{R}_{ij} \,\bar{J}_{i,-n} J_{j,-n} \right) \ket{\lambda, \bar{\lambda}} \label{ishi} \ee
We can now write down the most general boundary state preserving the symmetry. It takes the form of a linear sum of Ishibashi states, over the allowed charge sectors:
\be \ket{a; \mathcal{R}} = \sum_{\lambda \in \LambdaR} a_\lambda \, \kket{\lambda, -\mathcal{R} \lambda; \mathcal{R}} \label{almost}\ee
The Sugawara construction then ensures that since the state preserves each $U(1)_\alpha$, it also has no net energy inflow. Ishibashi states of the form \eqn{ishi} were also considered in \cite{ryu2,ryu1,janet,janet2}. It remains only to determine the complex coefficients $a_\lambda$.

\subsection{Clustering and the Cardy Condition} \label{cardysec}

The coefficients $a_\lambda$ in \eqn{almost} are constrained by two sets of consistency conditions. The first of these conditions is the requirement that correlation functions obey clustering. In this context, these are known as the {\it Cardy-Lewellen sewing conditions} \cite{cardyl, lewellen}. A nice review can be found in \cite{matthias}, with applications in \cite{gr1,gr2}. As imposing these sewing conditions is somewhat intricate, we relegate the details to Appendix \ref{sewingsec} where we show that the ratios of the coefficients $a_\lambda$ must obey
\be \frac{a_\lambda}{a_0} = e^{i \gamma_\mathcal{R}(\lambda)} \, e^{i \theta \cdot \lambda} \label{sewnup}\ee
This ratio is a phase, but with various parameters that we are free to choose. In particular, there are $N$ phases $\theta_i$. These are the generalisation of the phases that we met in \eqn{v} and \eqn{a}.

\para
The ratio \eqn{sewnup} also includes the factor $e^{i\gamma_\mathcal{R}(\lambda)}$. The definition of this phase is explained in Appendix \ref{sewingsec}. It does not play a role in many of the physical results that we derive below. For this reason, we do not elaborate on it any further in the main text. 

\para
While clustering imposes constraints on the ratios of the coefficients $a_\lambda$, it does not determine the overall normalisation. The upshot is that we are left with a family of boundary states, depending on the phases $\theta_i$, which preserve the symmetry $\mathcal{R}$ and take the form
\be \ket{\theta; \mathcal{R}} = g_\mathcal{R} \sum_{\lambda \in \LambdaR} e^{i \gamma_\mathcal{R}(\lambda)} \, e^{i \theta \cdot \lambda} \, \kket{\lambda, -\mathcal{R} \lambda; \mathcal{R}} \label{theta}\ee
We have taken the opportunity to rebrand the overall normalisation as $g_\mathcal{R} \equiv a_0$. This is appropriate, for $g_\mathcal{R}$ can be identified as the Affleck-Ludwig central charge of our boundary states \cite{afflud},
\be g_\mathcal{R} = \braket{0, 0}{\theta; \mathcal{R}}\nn\ee
This boundary central charge has a number of avatars; it can be thought of as the boundary contribution to the free energy $\mathcal{Z}_{AB}(q)$ or, relatedly, to the boundary entropy. For the boundary states \eqn{theta}, we claim that the correct normalisation is
\be g_\mathcal{R} = \sqrt{\vol(\LambdaR)} \label{g}\ee
where $\vol(\LambdaR)$ is the volume of the primitive unit cell of the lattice $\Lambda$. The boundary central charge has the property that $g_\mathcal{R} \geq 1$ but, as is to be expected, $g_\mathcal{R}$ need not be an integer. The same result for the central charge, albeit in a rather different setting, was previously derived in \cite{bachas} where it appeared as the tension of a D-brane.

\para
The normalisation $g_\mathcal{R}$ is fixed by the Cardy condition \eqn{cardy}. This states that the matrix element $\mathcal{Z}_P(q)$ computed between any two boundary states must have the interpretation of a partition function on an interval. For a general conformal field theory, this is the requirement that the partition function $\mathcal{Z}_{AB}(q)$ can be written as the sum of Virasoro characters in the open-string picture, weighted by positive integers.

\para
For us, there are two parts to the story. In this section, we will consider the Cardy condition with the same symmetry $\mathcal{R}$ imposed at the two ends of the interval. In this case the whole system has an unbroken $U(1)^N$ symmetry and the the Virasoro characters should be replaced by those of the appropriate chiral algebra. We will show that the normalisation \eqn{g} is the minimal choice that satisfies the Cardy condition. Applications of this condition can be found, for example, in \cite{sagi, green}.

\para
Ultimately, however, the Cardy condition is a statement about different boundary conditions $A$ and $B$ on each end of the interval, so we should study the system with two different symmetries $\mathcal{R}$ and $\mathcal{R}'$. We will turn to this in Section \ref{gsec} and show that the result \eqn{g} continues to hold.

\para
To proceed, we construct the Virasoro generators through the usual Sugawara construction,
\be
L_n = \frac{1}{2} \sum_{i=1}^N \sum_{m =\infty}^\infty \, {:} J_{i,m} J_{i,n-m} {:} \qquad
\bar{L}_n = \frac{1}{2} \sum_{i=1}^N \sum_{m =\infty}^\infty \, {:} \bar{J}_{i,m} \bar{J}_{i,n-m} {:} \label{sugawara}
\ee
The matrix element between two states, $\ket{\theta; \mathcal{R}}$ and $\ket{\theta'; \mathcal{R}}$, each of which preserves the same symmetry, is
\be \mathcal{Z}_P(q) = \bra{\theta'; \mathcal{R}} (-1)^F q^{\frac{1}{2}(L_0 + \bar{L}_0 - c/12)} \ket{\theta; \mathcal{R}} \nn\ee
where, for us, the bulk central charge is $c=N$. The factor of $(-1)^F$ is present because, if $\ket{A}$ describes some boundary condition, then the \emph{same} boundary condition at the other end is described by $\bra{A} \, (-1)^F$ rather than $\bra{A}$\footnote{This can be seen, for example, by computing the partition function of a single Dirac fermion. If $\ket{A}$ corresponds to the vector-like boundary condition \eqn{v} given by $\psi = e^{i\theta} \bar{\psi}$, then $\bra{A}$ corresponds to $\psi = -e^{i\theta} \bar{\psi}$. The need for this minus sign was also discussed in \cite{ryu2} (see footnote 69).}. Here $F$ is the holomorphic fermion number and should not be confused with the total fermion number $F + \bar{F}$.

\para
It might seem odd that we had to single out $F$ over $\bar{F}$. But there is actually no arbitrariness, as $F = \bar{F}$ holds for any valid boundary state. To see that this holds for our states $\ket{\theta; \mathcal{R}}$, note that acting on the ground state \eqn{lambda} in each charge sector $(\lambda, \bar{\lambda})$, the holomorphic fermion parity is given by
\be (-1)^F \ket{\lambda, \bar{\lambda}} = (-1)^{\sum_i \lambda_i} \ket{\lambda, \bar{\lambda}} = (-1)^{\lambda^2}\ket{\lambda, \bar{\lambda}} \nn\ee
where $\lambda^2 = \sum_i \lambda_i^2$. Similarly, the antiholomorphic fermion number is $(-1)^{\bar{F}} = (-1)^{\bar{\lambda}^2}$. But since we restrict to charge sectors obeying $\bar{\lambda} = -\mathcal{R} \lambda$, we necessarily have $\lambda^2 = \bar{\lambda}^2$ and so $F = \bar{F}$, as is necessary for a fermion in the presence of a boundary.

\para
With the same matrices $\mathcal{R}$ specifying both boundary states, the $\mathcal{R}$-dependence in the exponent of \eqn{ishi} cancels when taking the inner product. (This uses the fact that $\mathcal{R}^T \mathcal{R} = 1$.) Instead, the $\mathcal{R}$-dependence manifests itself only in the choice of lattice $\LambdaR$ that we sum over, with the matrix element given by
\be
\mathcal{Z}_P(q) &= g_\mathcal{R}^2 \sum_{\lambda \in \LambdaR} e^{i(\theta - \theta') \cdot \lambda} (-1)^{\lambda^2} q^{\frac{1}{4}(\lambda^2 + \bar{\lambda}^2)} \prod_{n=1}^\infty \frac{q^{-N/24}}{(1 - q^n)^N} \nn\\
&= g_\mathcal{R}^2 \sum_{\lambda \in \LambdaR} e^{i(\theta - \theta') \cdot \lambda} (-1)^{\lambda^2} \frac{q^{\frac{1}{2} \lambda^2}}{\eta(\tau)^N}
\nn\ee
where, in the Dedekind eta function, we've reverted to the argument $\tau$, related to $q$ via $q = e^{2\pi i \tau}$. The modular $\mathcal{S}$-transform of this partition function is
\be \mathcal{Z}_{AB}(q) = \int \! d^N x \left( g_\mathcal{R}^2 \sum_{\lambda \in \LambdaR} e^{i(\theta - \theta') \cdot \lambda} (-1)^{\lambda^2} e^{2\pi i x \cdot \lambda} \right) \frac{q^{\frac{1}{2} x^2}}{\eta(\tau)^N} \label{afters}\ee
In order that \eqn{afters} can be interpreted as an interval partition function of the form $\Tr_{\mathcal{H}_{AB}}(q^{L_0 - \frac{c}{24}})$, it must be a sum of Virasoro characters weighted by positive-integer coefficients. Actually, since both boundary conditions preserve the same $U(1)^N$ symmetry, these characters must fit together into representations of the corresponding chiral algebra,
\be [\Icurr_{\alpha,n}, \Icurr_{\beta,m}] = n \delta_{n+m,0} \mathcal{M}_{\alpha \beta} \nn\ee
where we've introduced $\mathcal{M}_{\alpha \beta} = Q_{\alpha i} Q_{\beta i} = \bar{Q}_{\alpha i} \bar{Q}_{\beta i}$. Irreducible representations of this algebra are labelled by common eigenvalues of $\Icurr_{\alpha, 0}$. We denote these eigenvalues as $\uplambda_\alpha$, by analogy with \eqn{lambda}. The Sugawara construction \eqn{sugawara} tells us that the Virasoro character associated to such an irrep is
\be {q^{\frac{1}{2} \uplambda^T \mathcal{M}^{-1} \uplambda}} \, \frac{1}{\eta(\tau)^N} \nn\ee
Since $\mathcal{M}$ is positive-definite, the power of $q$ is $\geq 0$. This means that the partition function \eqn{afters} must be the sum of terms $q^h \, \eta(\tau)^{-N}$ with $h \geq 0$, weighted by positive integers. Note that any real numbers $h$ are acceptable, because in general, the $\uplambda_\alpha$ need not obey any quantisation condition in the open sector.

\para
The above requirement is easily seen to hold. First write $(-1)^{\lambda^2} = e^{i \pi \sum_{i=1}^N \lambda_i} = e^{i \pi \cdot \lambda}$ in the integrand of \eqn{afters}. Then we can apply the standard identity
\be \sum_{\lambda \in \LambdaR} e^{2\pi i y \cdot \lambda} = \frac{1}{\vol(\LambdaR)}\sum_{\mu \in \LambdaR^\star} \delta^N(y - \mu) \nn\ee
where $\LambdaR^\star$ is the dual lattice, defined by the condition that $\mu \cdot \lambda \in \mathbb{Z}$ for all $\mu \in \LambdaR^\star$ and $\lambda \in \LambdaR$. The choice of $g_\mathcal{R}$ in \eqn{g} was designed to cancel the $1/\vol(\LambdaR)$ factor that arises in this sum. The upshot is that the partition function \eqn{afters} becomes
\be \mathcal{Z}_{AB}(q) = \sum_{\mu \in \LambdaR^\star} q^{\frac{1}{2}(\mu + \frac{\theta - \theta'}{2\pi} + \frac{1}{2})^2} \nn\ee
which is of the form promised.

\section{Boundaries Preserving Different Symmetries} \label{gsec}

The consistency conditions of the previous section resulted in a natural guess for a large family of boundary states $\ket{\theta; \mathcal{R}}$,
\be \ket{\theta; \mathcal{R}} = \sqrt{\vol(\LambdaR)} \sum_{\lambda \in \LambdaR} e^{i\gamma_\mathcal{R}(\lambda)} \, e^{i \theta \cdot \lambda} \, \kket{\lambda, -\mathcal{R} \lambda; \mathcal{R}} \nn\ee
However, the argument of the previous section does not fix the normalisation completely. For example, one could pick a positive integer $n_\mathcal{R}$ for each $\mathcal{R}$, and multiply each state by $\sqrt{n_\mathcal{R}}$, and they would continue to satisfy all the conditions we have imposed so far.

\para
One can demonstrate that for simple boundary conditions like those considered in the introduction, no such rescaling is necessary: the boundary states $\ket{\theta; \mathcal{R}}$ already reproduce the correct partition functions, computed via canonical quantisation. However, for more general boundary states which cannot be realised as linear boundary conditions on fermion fields, checking the normalisation this way is not an option.

\para
The first goal of this section is to show that the whole family of boundary states $\ket{\theta; \mathcal{R}}$ are, in fact, correctly normalised. To do this, we will check Cardy's condition between boundary states preserving different symmetries. We find that the partition function $\mathcal{Z}_{AB}$ is indeed always sensible, and that this comes about in a non-trivial way. The simplest interpretation is that all the integers $n_\mathcal{R}$ should be chosen to be 1.

\para
To start, we consider an interval in which different $U(1)^N$ symmetries are preserved at each end. The associated symmetries are those described by $\mathcal{R}$ and $\mathcal{R}'$ respectively, and the matrix element is
\be \mathcal{Z}_P(q) = \bra{\theta'; \mathcal{R}'} (-1)^F q^{\frac{1}{2} (L_0 + \bar{L}_0 - c/12)}\ket{\theta; \mathcal{R}} \label{z}\ee
This time, the $\mathcal{R}$ matrices in the exponent \eqn{ishi} of the two states do not cancel. A direct evaluation gives
\be
\mathcal{Z}_P(q) = g_\mathcal{R} g_{\mathcal{R}'}
\left[ \sum_{\lambda \in \LambdaRRdash} e^{i (\gamma_\mathcal{R}(\lambda) - \gamma_{\mathcal{R}'}(\lambda))} \, e^{2\pi i (\frac{\theta - \theta'}{2\pi} + \frac{1}{2}) \cdot \lambda} \, q^{\frac{1}{2} \lambda^2} \right]
\! \frac{1}{q^{N/24}} \,
\prod_{n=1}^\infty \frac{1}{\det \! \left( \mathds{1} - q^n \mathcal{R}^T \mathcal{R}' \right)}
\nn\ee
Here we have introduced a new lattice $\LambdaRRdash$, which arises from the need to sum over only those charge sectors $(\lambda, \bar{\lambda})$ compatible with both symmetries---that is, satisfying both $\bar{\lambda} = -\mathcal{R} \lambda$ and $\bar{\lambda} = -\mathcal{R}' \lambda$. For these reasons, we shall call it the `intersection lattice'. It is defined by
\be \LambdaRRdash = \Big\{\, \lambda \in \mathbb{Z}^N \ : \ \mathcal{R} \lambda = \mathcal{R}' \lambda \in \mathbb{Z}^N \ \Big\} \label{lambdaintersect}\ee
We would like to compute the transformation of the partition function $\mathcal{Z}_P(q)$ under the modular $\mathcal{S}$-transformation. We start by dealing with the factor
\be q^{N/24} \prod_{n=1}^\infty \det \left( 1 - q^n \mathcal{R}^T \mathcal{R}' \right) = \prod_r q^{1/24} \prod_{n=1}^\infty \left( 1 - r q^n \right) \nn\ee
where the product $\prod_r$ is over the $N$ eigenvalues of $\mathcal{R}^T \mathcal{R}'$. Since this is an orthogonal matrix, its eigenvalues are either $\pm 1$ or occur in complex-conjugate pairs of phases. To establish notation for this, we introduce
\be n_\pm = \text{Number of $\pm 1$ eigenvalues} \nn\ee
We then write the remainder as $e^{\pm 2\pi it}$, where $t$ ranges over some multiset $T \subset (0, \frac{1}{2})$. The various contributions of these eigenvalues to the product are
\be
{+1} &\quad\Rightarrow\quad q^{1/24} \prod_{n=1}^\infty (1 - q^n) = \eta(\tau) \nn\\
{-1} &\quad\Rightarrow\quad q^{1/24} \prod_{n=1}^\infty (1 + q^n) = \frac{\eta(2\tau)}{\eta(\tau)} \nn\\
{e^{\pm 2\pi it}} &\quad\Rightarrow\quad q^{1/12} \prod_{n=1}^\infty \left( 1 - e^{2\pi it} q^n \right) \! \left( 1 - e^{-2\pi it} q^n \right) = \frac{1}{2 \sin(\pi t)} \frac{\theta_1 (t|\tau)}{\eta(\tau)}
\nn\ee
where we've adopted the theta-function conventions of \cite{yellow}. For each of these, the modular $\mathcal{S}$-transformations are given by
\be
\eta(\tau) &\quad\longrightarrow\quad \sqrt{-i \tau} \, \eta(\tau) \nn\\
\frac{\eta(2\tau)}{\eta(\tau)} &\quad\longrightarrow\quad \frac{1}{\sqrt{2}} \frac{\eta(\tau / 2)}{\eta(\tau)} \nn\\
\frac{1}{2\sin(\pi t)} \frac{\theta_1(t|\tau)}{\eta(\tau)} &\quad\longrightarrow\quad -\frac{i \, q^{t^2 / 2}}{2\sin(\pi t)} \frac{\theta_1(t\tau|\tau)}{\eta(\tau)}
\nn\ee

\para
Next, we deal with the factor in $\mathcal{Z}_P(q)$ involving the sum over lattice sites. We need to write the factor of $e^{i (\gamma_\mathcal{R}(\lambda) - \gamma_{\mathcal{R}'}(\lambda))}$ as an exponential linear in $\lambda$. For this, we appeal to a fact from Appendix~\ref{sewingsec}, which states that for all $\lambda \in \LambdaRRdash$,
\be e^{i (\gamma_\mathcal{R}(\lambda) - \gamma_{\mathcal{R}'}(\lambda))} = (-1)^{s \cdot \lambda} \nn\ee
for some vector $s \in \LambdaRRdash^\star$. The exact expression for $s$ won't concern us here. With the sum now in the form of a theta function, we can proceed as before, this time using the modular $\mathcal{S}$-transformation property
\be
\sum_{\lambda \in \LambdaRRdash} e^{2\pi i y \cdot \lambda} q^{\frac{1}{2} \lambda^2}
\quad\longrightarrow\quad
\sqrt{-i\tau} \,^{\dim(\LambdaRRdash)} \frac{1}{\vol(\LambdaRRdash)} \sum_{\mu \in \LambdaRRdash^\star} q^{\frac{1}{2} (\mu + \Pi(y))^2}
\nn\ee
where $\Pi(y)$ denotes the orthogonal projection of the vector $y$ onto the subspace spanned by $\LambdaRRdash$. Combining everything so far, we have
\be
\mathcal{Z}_{AB}(q) = {}& g_\mathcal{R} g_{\mathcal{R}'}
\bigg[ \frac{\sqrt{-i\tau} \,^{\dim(\LambdaRRdash)}}{\vol(\LambdaRRdash)} \sum_{\mu \in \LambdaRRdash^\star} q^{\frac{1}{2} \vstretch{1.2}{(} \mu + \Pi (\frac{s}{2} + \frac{\theta-\theta'}{2\pi} + \frac{1}{2}) \vstretch{1.2}{)}^2} \bigg] \nn\\
& \ \ \ \ \ \ \ \ \times
\bigg[ \frac{1}{\sqrt{-i\tau} \, \eta(\tau)} \bigg]^{n_+} \;
\bigg[ \frac{\sqrt{2} \, \eta(\tau)}{\eta(\tau / 2)} \bigg]^{n_-} \,
\prod_{t \in T} \bigg[ \frac{2i \sin(\pi t)}{q^{t^2 / 2}} \frac{\eta(\tau)}{\theta_1(t\tau|\tau)} \bigg]
\nn\ee
Importantly, factors of $\sqrt{-i\tau}$ appear in two places: there are $\dim(\LambdaRRdash)$ factors from the lattice factor, and $-n_+$ from the $+1$ eigenvalues of $\mathcal{R}^T\mathcal{R}'$. If we are to interpret this as the partition function of a theory on the interval, these must cancel meaning that we must have $\dim(\LambdaRRdash) = n_+$. Happily this is the case, as can be seen from the definition \eqn{lambdaintersect}, which says that $\lambda$ is constrained to obey $\mathcal{R}^T \mathcal{R}' \lambda = \lambda$.

\para
Another immediate simplification is to make the replacement
\be (\! \sqrt{2} \,)^{n_-} \prod_{t \in T} 2\sin(\pi t) = \sqrt{\detprime(\mathds{1} - \mathcal{R}^T \mathcal{R}')} \nn\ee
where $\detprime$ denotes the product over non-zero eigenvalues.

\para
The upshot is that the $\mathcal{S}$-transformed partition function is given by
\be
\mathcal{Z}_{AB}(q) = {}& \frac{\sqrt{\vol(\LambdaR) \, \vol(\LambdaRdash)}}{\vol(\LambdaRRdash)} \sqrt{\detprime(\mathds{1} - \mathcal{R}^T \mathcal{R}')} \nn\\
& \ \ \times \sum_{\mu \in \LambdaRRdash^\star} q^{\frac{1}{2} \vstretch{1.2}{(} \mu + \Pi(\frac{s}{2} + \frac{\theta-\theta'}{2\pi} + \frac{1}{2}) \vstretch{1.2}{)}^2} \label{itscomplicated}\\
& \ \ \ \ \ \ \ \times \bigg[ \frac{\eta(\tau)}{\eta(\tau / 2)} \bigg]^{n_-}
\prod_{t \in T} \bigg[ \frac{1}{q^{t^2 / 2}} \frac{i \, \eta(\tau)}{\theta_1(t\tau|\tau)} \bigg]
\, \frac{1}{\eta(\tau)^{n_+}} \nn
\ee
We have separated the terms into three groups, each of which will play its own distinct role in what follows.

\subsection{Ground State Degeneracy}

The partition function \eqn{itscomplicated} describes the fermions on the interval, with different boundary conditions on the left and right, corresponding to $\ket{\theta'; \mathcal{R}'}$ and $\ket{\theta; \mathcal{R}}$ respectively. We would like to compute the number of ground states of this system.

\para
Consider first the final term $1 / \eta(\tau)^{n_+}$. The integer $n_+$ has yet a third interpretation: the intersection of the two $U(1)^N$ symmetry groups preserved by the two boundaries $\mathcal{R}$ and $\mathcal{R}'$ is $U(1)^{n_+}$. To see this, note that a common $U(1)$ symmetry corresponds to a pair of vectors $s_\alpha \in \mathbb{Z}^N$ and $s'_\alpha \in \mathbb{Z}^N$ such that $(Q_{i \alpha}, \bar{Q}_{i \alpha}) s_\alpha = (Q'_{i \alpha}, \bar{Q}'_{i \alpha}) s'_\alpha$. In terms of the vector $Q_{i \alpha} s_\alpha$, these conditions again reduce to the requirement that $Q_{i \alpha} s_\alpha$ is an eigenvector of $\mathcal{R}^T \mathcal{R}'$ with eigenvalue $+1$.

\para
We can then run a similar argument to what we saw in Section \ref{cardysec}: because the boundary conditions preserve a common $U(1)^{n_+}$, the Hilbert space must furnish a representation of the $\mathfrak{u}(1)^{n_+}$ current algebra. The structure of such representations forces the partition function to contain a factor of $1 / \eta(\tau)^{n_+}$. Thus, the final term of \eqn{itscomplicated} is necessarily present in order that the partition function be valid, but as far as the degeneracy is concerned, it can be discarded.

\para
Other terms of \eqn{itscomplicated} have no bearing on either the validity of the partition function or the degeneracy. In particular, for these purposes we can completely ignore
\be
\frac{\eta(\tau)}{\eta(\tau / 2)} &= q^{1/48} \prod_{n=1}^\infty (1 + q^{n/2}) \nn\\
\frac{i \, \eta(\tau)}{\theta_1(t\tau|\tau)} &= q^{-1/12} q^{t/2} \prod_{n=0}^\infty \left( 1 - q^{n+t} \right)^{-1} \left(1 - q^{n+1-t} \right)^{-1}
\nn\ee
as both are power series with positive integer coefficients and leading coefficient unity.\footnote{Both of these factors also supply a factor of $\frac{1}{\eta(\tau)}$. For the first, this follows from the identity $\frac{\eta(\tau)}{\eta(\tau / 2)} = \frac{1}{\eta(\tau)} \sum_{n=0}^\infty q^{(n+1/2)^2 / 4}$. For the second, such a representation is also possible, although not in simple closed form. So \eqn{itscomplicated} actually contains many more than $n_+$ copies of $\frac{1}{\eta(\tau)}$.} Using these expressions, one can also check that all powers of $q$ occurring in \eqn{itscomplicated} have exponent $\geq -N/24$. That is, all Virasoro weights in the open sector are $\geq 0$, as is consistent for a unitary theory.

\para
The lattice term
\be \sum_{\mu \in \LambdaRRdash^\star} q^{\frac{1}{2} \vstretch{1.2}{(} \mu + \Pi(\frac{s}{2} + \frac{\theta-\theta'}{2\pi} + \frac{1}{2}) \vstretch{1.2}{)}^2} \label{spinme}\ee
is more interesting. For generic values of the phases, parameterised by $\theta$ and $\theta'$, this power series has leading coefficient unity. However, at certain symmetrical values of the phases, the coefficient of the leading term may jump from 1 to a higher value. This corresponds to the kind of behaviour we saw in the introduction, where the ground state degeneracy of a single Dirac fermion on an interval is typically 1, but may jump to 2 when the boundary state phases align.

\para
Not all the phases affect the physics. Rather, only the orthogonal projection of $\theta - \theta'$ onto $\LambdaRRdash$, which can naturally be thought of as living in $\text{Hom}(\LambdaRRdash, U(1)) \cong U(1)^{n_+}$, enters into the exponent of \eqn{spinme}. This implies that the less compatible the boundary conditions, the fewer means we have to affect them. This mirrors what we saw in the introduction for a single Dirac fermion.

\para
In what follows, we first assume generic values of the phases so that \eqn{spinme} has no degeneracy. Later, when we discuss specific examples, we will explore how the ground state degeneracy jumps at specific values of the phases.

\para
After stripping off all of the terms discussed so far, what's left of the partition function determines the ground state degeneracy. It is given by
\be G[\mathcal{R},\mathcal{R}'] = \frac{\sqrt{\vol(\LambdaR) \, \vol(\LambdaRdash)}}{\vol(\LambdaRRdash)} \sqrt{\detprime(\mathds{1} - \mathcal{R}^T \mathcal{R}')} \label{G}\ee
As a sanity check, note that if we put the same boundary conditions on each end, then we generically have a unique ground state: $G[\mathcal{R}, \mathcal{R}] = 1$. We will give a number of more intricate examples in Section \ref{examplesec}. This formula bears a tantalising similarity to a result by Kapustin on the ground state degeneracy of Abelian quantum Hall states with topological order on the boundary \cite{kapustin}; it would be interesting to understand this relation better. 

\para
The number of ground states of the system should be an integer. Indeed, this is one of the key requirements of the boundary conformal field theory approach. It is not at all obvious that $G[\mathcal{R}, \mathcal{R}']$, defined in \eqn{G}, is integer-valued. We claim that it almost is.

\para
Specifically, we show that -- under certain circumstances that we detail below -- the matrices $\mathcal{R}$ fall into two separate classes which, following the introduction, we call vector-like $\classv$ and axial-like $\classa$. When $\mathcal{R}$ and $\mathcal{R}'$ are both taken from the same class, the ground state degeneracy is indeed an integer as it should be. However, if $\mathcal{R} \in \classv$ and $\mathcal{R}' \in \classa$, we find $G[\mathcal{R}, \mathcal{R}'] \in \sqrt{2} \, \mathbb{Z}$. The interpretation of this is that the two classes of ground states are mutually incompatible since they give rise to a Majorana zero mode.

\subsection{The Two Classes of Boundary States} \label{twoclasses}

We conjecture that $G[\mathcal{R}, \mathcal{R}']$ takes values in $\mathbb{Z} \cup \sqrt{2} \, \mathbb{Z}$. Further, we conjecture the existence of two classes $\classv$ and $\classa$ such that the presence of a $\sqrt{2}$ is dictated by whether $\mathcal{R}$ and $\mathcal{R}'$ lie in different classes.

\para
These conjectures do not seem easy to prove in full generality. We have been able to demonstrate that they hold in large classes of examples. In this section, we will show the following.
\begin{itemize}
\item Task 1: For a large class of examples, we prove the above conjectures, and, in the process, extract a criterion that determines which of the two classes $\classv$ and $\classa$ a given symmetry $\mathcal{R}$ falls into.
\item Task 2: For an even larger class of examples, we prove a weaker version with $\mathbb{Z} \cup \sqrt{2} \, \mathbb{Z}$ replaced with $\mathbb{Q} \cup \sqrt{2} \, \mathbb{Q}$, again extracting a criterion for the classes $\classv$ and $\classa$.
\item Task 3: By assuming the conjecture holds, we obtain a concrete criterion for the classes $\classv$ and $\classa$ in the general case.
\end{itemize}
Furthermore, in randomised numerical experiments, it is found that in all cases, the classes $\classv, \classa$ derived in the third line correctly predict whether $G[\mathcal{R}, \mathcal{R}']$ lies in $\mathbb{Z}$ or $\sqrt{2} \, \mathbb{Z}$, with no other values possible. We feel that this is convincing evidence in favour of the conjectures.

\subsubsection{Task 1}

In this section, we limit ourselves to choices of $\mathcal{R}$ and $\mathcal{R}'$ obeying the following two properties:
\begin{enumerate}[i)]
\item $\LambdaRRdash = \{0\}$ or, equivalently, $n_+=0$. This ensures that $\mathcal{R} - \mathcal{R}'$ is non-singular. Under this assumption, the number of ground states \eqn{G} takes the simplified form
\be G[\mathcal{R},\mathcal{R}'] = \sqrt{\vol(\LambdaR) \, \vol(\LambdaRdash) \, |\!\det(\mathcal{R}-\mathcal{R}')|} \label{Gsimple}\ee
\item Neither $\mathcal{R}$ nor $\mathcal{R}'$ have eigenvalue $-1$. This allows the Cayley parameterisations
\be \mathcal{R} = \frac{\mathds{1} - A}{\mathds{1} + A} \ \ \ \text{and} \ \ \ \mathcal{R}' = \frac{\mathds{1}- A'}{\mathds{1} + A'} \nn\ee
This gives a one-to-one correspondence between the rational orthogonal matrix $\mathcal{R}$ with no $-1$ eigenvalues, and the rational anti-symmetric matrix $A$. The ground state degeneracy can then be written as
\be G[\mathcal{R}, \mathcal{R}'] = 2^{N/2} \, |\text{Pf}(A - A')| \, \sqrt{\frac{\vol(\LambdaR) \, \vol(\LambdaRdash)}{\det(\mathds{1} + A)\det(\mathds{1}+A')}}\nn\ee
\end{enumerate}
Note that the combined requirements of i) and ii) mean that this proof holds only for rotation matrices with $N$ even, but other than that, these assumptions are generic.

\paragraph{A simple warm-up} \mbox{} \\
To begin with, we add one more assumption, namely that $A$, $A'$ are integer-valued rather than merely rational-valued. This is straightforward to relax, and we will do so shortly. With these assumptions in place, we now associate an integer $n \in \{0,\ldots,N\}$ to the matrix $\mathcal{R}$,
\be n = \text{nullity}_{\,\mathbb{F}_2}(\mathds{1} + A) \nn\ee
That is, $n$ is the dimension of the kernel of the $N\times N$ matrix $\mathds{1} + A$, regarded over the finite field $\mathbb{F}_2$. (Equivalently, $n$ is the number of linearly independent vectors, with integer elements defined mod 2, which map to even-integer vectors under $\mathds{1} + A$.) We then have, as shown in Appendix~\ref{latticesappendix},
\be \vol(\LambdaR) = 2^{-n} \det(\mathds{1} + A) \nn\ee
Similarly, we can define the integer $n'$ associated to $\mathcal{R}'$. The ground state degeneracy can then be written as
\be G[\mathcal{R}, \mathcal{R}'] = |\text{Pf}(A - A')| \, (\!\sqrt{2}\,)^{N-n-n'} \nn\ee
This is sufficient to prove the result we want, provided that $N \geq n+n'$. However, if $N < n+n'$ then we seemingly have a negative power of $\sqrt{2}$ and have to work a little harder. In fact, this situation has a nice linear-algebraic interpretation, since it guarantees that the two kernels intersect,
\be \dim_{\,\mathbb{F}_2} \! \Big( \ker_{\,\mathbb{F}_2} (\mathds{1} + A) \cap \ker_{\,\mathbb{F}_2} (\mathds{1} + A') \Big) \geq n+n'-N \nn\ee
That certainly implies
\be \text{nullity}_{\,\mathbb{F}_2}(A - A') \geq n+n'-N \label{itsnull}\ee
We now utilise the fact that $A-A'$, being an antisymmetric integer matrix, has a Smith-like normal form,
\be U (A - A') U^T = \bigoplus_{i=1}^{N/2} \left(\begin{array}{cc}0 & \nu_i \\ -\nu_i & 0\end{array}\right) \nn\ee
where $U$ is unimodular and the $\nu_i$ are integers. The nullity in equation \eqn{itsnull} is then given in terms of this data by
\be \text{nullity}_{\,\mathbb{F}_2}(A - A') = 2 \, \cdot \text{\#(even $\nu_i$)} \nn\ee
We can conclude that there are at least $\left\lceil (n+n'-N)/2 \right\rceil$ even $\nu_i$. Then, since
\be \text{Pf}(A - A') = \prod_{i=1}^{N/2} \nu_i \nn\ee
it follows that the pfaffian is divisible by $2^{\left\lceil \frac{1}{2}(n+n'-N)\right\rceil}$, which is just enough to offset the dangerous negative power of $(\!\sqrt{2}\,)^{\frac{1}{2}(N-n-n')}$. This ensures that, in all cases, $G[\mathcal{R},\mathcal{R}']$ is an integer, or an integer times $\sqrt{2}$, as promised.

\para
The derivation above also provides the criterion for whether a given boundary condition sits in class $\classv$ or class $\classa$. Since $N$ is even, the irrational part of $G[\mathcal{R},\mathcal{R}']$ is given by $(\!\sqrt{2}\,)^{n+n'}$. The ground state degeneracy fails to be an integer if $n \neq n'$ mod 2. In other words, the class of boundary condition $\mathcal{R}$ is determined by $n$ mod 2.

\paragraph{The rational case} \mbox{} \\
With a little extra work, we can re-run the arguments of the last section in the case where $A,A'$ are rational-valued. Once again, we start from
\be G[\mathcal{R}, \mathcal{R}'] = 2^{N/2} \, |\text{Pf}(A - A')| \, \sqrt{\frac{\vol(\LambdaR) \, \vol(\LambdaRdash)}{\det(\mathds{1} + A)\det(\mathds{1}+A')}}\nn\ee
The difference now is that the pfaffian may only be rational, and therefore its denominator has to emerge out of the second expression in order to cancel it. To see how this works, we first need to construct a bunch of auxiliary data associated to $\mathcal{R}$:
\begin{itemize}
\item Write $A = \tilde{A} / g$ where $\tilde{A}$ is an integer matrix.
\item Compute the Smith-like decomposition of $\tilde{A}$,
\be \tilde{A} = U \! D U^T \qquad D = J \, \text{ddiag}(\nu_i) \qquad J = \bigoplus_{i=1}^{N/2} \left(\begin{array}{cc}0 & 1\\-1 & 0\end{array}\right) \nn\ee
where by `$\text{ddiag}(\nu_{1},\nu_{2},\dots)$' we mean the diagonal matrix with each entry repeated twice, that is $\text{diag}(\nu_{1},\nu_{1},\nu_{2},\nu_{2},\dots)$.
\item Define integers $g_i = \gcd(g, \nu_i)$.
\item Define an integer matrix
\be X = U^{T,-1} \text{ddiag}(g/g_i) + UJ \, \text{ddiag}(\nu_i/g_i) \nn\ee
\end{itemize}
The analog of the integer $n$ from the previous section is then defined to be
\be n = \text{nullity}_{\,\mathbb{F}_2}(X) \nn\ee
It is shown in Appendix~\ref{latticesappendix} that
\be \vol(\LambdaR) = 2^{-n} \det(\mathds{1} + A)\, \prod_{i=1}^{N/2} (g/g_i)^2 \nn\ee
The new part of this expression is the product on the right. This will turn out to be precisely the integer required to cancel the denominator of the pfaffian. To see this, we plug the above result into the ground state degeneracy, yielding the result
\be G[\mathcal{R}, \mathcal{R}'] = \frac{|\text{Pf}(g'\!\tilde{A} - g \tilde{A}')|}{\prod_{i=1}^{N/2} g_i \, g_i'} \, (\!\sqrt{2}\,)^{N-n-n'} \label{Gsimple2}\ee
It's not hard to show that the fraction is integer-valued. For example, one can simply write it as
\be
\frac{|\text{Pf}(g' \! \tilde{A} - g \tilde{A}')|}{\prod_{i=1}^{N/2} g_i \, g_i'}
&=
\det\!\Big[ [UJ \text{ddiag}(\tfrac{\nu_i}{g_i})]^T [U^{\prime,T,-1} \text{ddiag}(\tfrac{g'}{g_i'})] \label{pfoverint}\\
& \qquad \ \ {} + [U^{T,-1} \text{ddiag}(\tfrac{g}{g_i})]^T [U'J \text{ddiag}(\tfrac{\nu_i'}{g_i'})] \Big]^{1/2} \nn
\ee
Since the matrix involved on the right is an integer-valued one, the right side is manifestly the square root of an integer. Unfortunately, it's no longer manifestly rational. However, the left side is, so putting the two pieces of information together shows that the whole thing is indeed an integer.

\para
The final piece of the argument is to show that the power of $\sqrt{2}$ in \eqn{Gsimple2}, if it ever goes negative, is compensated for by \eqn{pfoverint} becoming divisible by a power of 2. In the previous section, this went via an argument involving the intersection of two kernels. Similarly, here it follows from the linear-algebraic fact that for $N \times N$ matrices $A,B,A',B'$,
\be \text{nullity}(A^T B' - B^T A') \geq \text{nullity}(A - B) + \text{nullity}(A' - B') - N \nn\ee
Applied to our situation, this says that the matrix on the right hand side of \eqn{pfoverint} has $\mathbb{F}_2$-nullity at least $n+n'-N$, and therefore that its determinant is divisible by $2^{n+n'-N}$. Then, since \eqn{pfoverint} is an integer, it follows that \eqn{pfoverint} is divisible by $2^{\left\lceil \frac{1}{2}(n+n'-N)\right\rceil}$. This establishes the claimed integrality property of $G[\mathcal{R}, \mathcal{R}']$.

\subsubsection{Task 2}

In this section we will concern ourselves purely with the irrational part of $G[\mathcal{R}, \mathcal{R}']$. By freeing us of the burden of having to show that the rational part is actually an integer, we will be able to establish the rest of the conjecture in greater generality.

\para
This time we will only rely on the assumption that $\mathcal{R}$ and $\mathcal{R}'$ have a Cayley parametrisation. This assumption restricts us to rotation matrices, but is otherwise generic. We start from the general expression \eqn{G},
\be G[\mathcal{R},\mathcal{R}'] = \frac{\sqrt{\vol(\LambdaR) \, \vol(\LambdaRdash)}}{\vol(\LambdaRRdash)} \sqrt{\detprime(\mathds{1} - \mathcal{R}^T \mathcal{R}')} \nn\ee
The new ingredients are the volume of the intersection lattice, and the replacement of $\det$ with $\detprime$. As we shall see, these complications cancel one another. Let us deal with the latter complication first. Substituting in the Cayley parametrisations, we have
\be \detprime(\mathds{1} - \mathcal{R}^T \mathcal{R}') = \detprime \! \left( (-2) \frac{1}{(\mathds{1} - A)(\mathds{1} + A')}(A - A') \right) \nn\ee
In the previous sections, we could simply pull out the factors of $\frac{1}{\mathds{1} - A}$ and $\frac{1}{\mathds{1} - A'}$ from the determinant. However, for $\detprime$, this is no longer an allowed operation. Instead, we must invoke the Smith-like decomposition of $A - A'$,
\be
U (A - A') U^T = D = \bigoplus_{i=1}^{(N-k)/2}
\left(\begin{array}{cc}0 & \nu_i \\ -\nu_i & 0\end{array}\right) \oplus \bigoplus_{i=1}^k \left(0\right)
\nn\ee
where $U$ is unimodular, the $\nu_i$ are nonzero rationals, and $k$ is the nullity of $A - A'$. Inserting this decomposition into the previous expression, we may then separate it into two factors as follows:
\be
\detprime(\mathds{1} - \mathcal{R}^T \mathcal{R}') &= 2^{N-k} \, \detprime \! \left( \frac{1}{U(\mathds{1} - A)(\mathds{1} + A')U^T} \, D\right) \nn\\
&= 2^{N-k} \, \det \! \bigg( \frac{1}{U(\mathds{1} - A)(\mathds{1} + A')U^T} \bigg|_{\includegraphics[scale=.45]{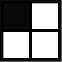}} \, \bigg) \, \detprime(D) \nn
\ee
Here, the symbol in front of the matrix in the second line instructs us to restrict to the top-left $(N-k) \times (N-k)$ block of that matrix. For the identity we have just used to be valid, this block must be invertible; one can check that this is indeed the case.

\para
We now shift our attention to the term $\vol(\LambdaRRdash)$. To deal with this, we need to find a parametrisation of the lattice $\LambdaRRdash$. Recalling definition \eqn{lambdaintersect},
\be \LambdaRRdash = \Big\{\, \lambda \in \mathbb{Z}^N \ : \ \mathcal{R} \lambda = \mathcal{R}' \lambda \in \mathbb{Z}^N \ \Big\} \nn\ee
we see that $\lambda$ is necessarily an element of $\ker(\mathcal{R} - \mathcal{R}')$. All elements of this kernel can be parametrised as $(\mathds{1} + A') U^T v$, where $v$ is a vector of the form $v = (0, \dots, 0, v_1, \dots, v_k)$, i.e. only its last $k$ components are nonzero. We will use $\mathbb{R}^k$ to denote such vectors. On top of this, $v$ is constrained by the fact that both $\lambda$ and $\mathcal{R}' \lambda$ must be integer vectors, which forces $v$ to lie in the sublattice
\be \Lambda_v = \Big\{\, v \in \mathbb{R}^k \ : \ (\mathds{1} + A') U^T v \in \mathbb{Z}^N \ , \ (\mathds{1} - A') U^T v \in \mathbb{Z}^N \ \Big\} \label{Lambdav}\ee
It follows that the lattice volume we are interested in is
\be \vol(\LambdaRRdash) = \det \! \Big( U (\mathds{1} - A) (\mathds{1} + A') U^T \Big|_{\includegraphics[scale=.45]{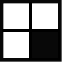}} \, \Big)^{1/2} \, \vol(\Lambda_v) \nn\ee
where this time, the symbol in front of the matrix instructs us to restrict to its lower-right $k \times k$ block.

\para
Let us return to the ground state degeneracy. Inserting the results so far, we have
\be G[\mathcal{R},\mathcal{R}'] = 2^{(N-k)/2} \, \frac{\sqrt{\vol(\LambdaR) \, \vol(\LambdaRdash)}}{\vol(\Lambda_v)} \sqrt{\frac{\det \! \big( (U(\mathds{1} - A)(\mathds{1} + A')U^T)^{-1} |_{\includegraphics[scale=.35]{boxtl.pdf}} \, \big)}{\det \! \big( U (\mathds{1} - A) (\mathds{1} + A') U^T |_{\includegraphics[scale=.35]{boxbr.pdf}} \, \big)}} \sqrt{ \detprime(D) } \nn\ee
As remarked at the start, we shall be content to focus only on the the irrational part of this expression. To this end, we may immediately drop certain factors. For example, the term
\be 2^{(N-k)/2} \nn\ee
is rational, since $N - k$ is an even number. So also is the term
\be \vol(\Lambda_v) \nn\ee
as $\Lambda_v$ is a rank-$k$ sublattice of $\mathbb{R}^k$, whose elements $v$ are defined by the conditions \eqn{Lambdav} that certain integer-linear combinations of their components $v_i$ are integers. Similarly,
\be \sqrt{\detprime(D)} = \prod_{i=1}^{(N-k)/2} \nu_i \nn\ee
where each of the $\nu_i$ is rational. Finally, we may invoke the linear-algebraic fact that
\be \sqrt{\frac{\det \! \big( (U(\mathds{1} - A)(\mathds{1} + A')U^T)^{-1} |_{\includegraphics[scale=.35]{boxtl.pdf}} \, \big)}{\det \! \big( U (\mathds{1} - A) (\mathds{1} + A') U^T |_{\includegraphics[scale=.35]{boxbr.pdf}} \, \big)}} = \frac{1}{\sqrt{\det(\mathds{1} + A) \, \det(\mathds{1} + A')}} \nn\ee
to rewrite the remaining irrational part as
\be G[\mathcal{R},\mathcal{R}']_\text{irrational} = \sqrt{\frac{\vol(\LambdaR) \, \vol(\LambdaRdash)}{\det(\mathds{1} + A) \, \det(\mathds{1} + A')}} \nn\ee
This is something we have calculated before. Indeed, when $N$ is even, we have already seen how to associate to a matrix $\mathcal{R}$ integers $n, g, g_i$ such that
\be \vol(\LambdaR) = 2^{-n} \det(\mathds{1} + A)\, \prod_{i=1}^{N/2} (g/g_i)^2 \nn\ee
Here we need the extension of this result to matrices with odd $N$. As before, one first constructs a set of auxiliary data:
\begin{itemize}
\item Write $A = \tilde{A} / g$ where $\tilde{A}$ is an integer matrix.
\item Compute the Smith-like decomposition of $\tilde{A}$,
\be \tilde{A} = U \! D U^T \qquad D = [J \, \text{ddiag}(\nu_i)] \oplus (0) \qquad J = \bigoplus_{i=1}^{(N-1)/2} \left(\begin{array}{cc}0 & 1\\-1 & 0\end{array}\right) \nn\ee
\item Define integers $g_i = \gcd(g, \nu_i)$.
\item Define an integer matrix
\be X = U^{T,-1} [\text{ddiag}(g/g_i) \oplus (1)] + UJ \, [\text{ddiag}(\nu_i/g_i) \oplus (0)] \nn\ee
\end{itemize}
The integer $n$ associated to $\mathcal{R}$ is then $n = \text{nullity}_{\, \mathbb{F}_2}(X)$. The analogous result for the lattice volume is
\be \vol(\LambdaR) = 2^{-n} \det(\mathds{1} + A)\, \prod_{i=1}^{(N-1)/2} (g/g_i)^2 \nn\ee
The upshot of all this is that the irrational part of the ground state degeneracy is simply given by
\be G[\mathcal{R},\mathcal{R}']_\text{irrational} = (\!\sqrt{2}\,)^{n + n'} \nn\ee
We thus conclude, as before, that the class of $\mathcal{R}$ is dictated by the value of $n \text{ mod } 2$.

\subsubsection{Task 3}

In the last section, we saw how to associate an integer $n \in \{0, \dots, N\}$ to any matrix $\mathcal{R}$ that admits a Cayley parametrisation, such that the two classes of boundary states are labelled by $n \text{ mod } 2$.

\para
Here, we would like to cast away the final crutch of the existence of a Cayley parametrisation. To do this, we appeal to a classical result of Liebeck-Osborne \cite{liebeckosborne}, which states that any rational orthogonal matrix $\mathcal{R}$ can be written as
\be \mathcal{R} = D \, \mathcal{R}_0 \nn\ee
where $D = \text{diag}(\pm 1, \dots , \pm 1)$, and $\mathcal{R}_0$ admits a Cayley parametrisation. It is not hard to show that the irrational part of $G[\mathcal{R}, \mathcal{R}_0]$ is simply
\be (\!\sqrt{2}\,)^{n_-} \nn\ee
where $n_-$ is the number of negative eigenvalues of $D$. This suggests the following definition of the integer $n(\mathcal{R})$ for a general matrix $\mathcal{R}$. First write $\mathcal{R} = D \, \mathcal{R}_0$ in the form above. Then set
\be n(\mathcal{R}) = n(\mathcal{R}_0) + n_- \mod 2 \label{ngeneral}\ee
where $n(\mathcal{R}_0)$ is calculated as in the previous section. As discussed at the start, numerical experiments then suggest that the conjecture
\be
G[\mathcal{R}, \mathcal{R}'] \in \begin{cases}\mathbb{Z} & n(\mathcal{R}) = n(\mathcal{R}') \\ \mathbb{Z} \sqrt{2} & n(\mathcal{R}) \neq n(\mathcal{R}')\end{cases}
\ee
continues to hold even in the cases that remain unaddressed by our proof.

\paragraph{Properties of ${n(\mathcal{R})}$} \mbox{} \\
It is natural to ask whether the map that we have defined,
\be n : O(N, \mathbb{Q}) \longrightarrow \mathbb{Z}_2 \nn\ee
is a group homomorphism. Or perhaps the opposite quantity, $1 - n$? It turns out that for general $N$, both statements are false. However, as we shall see in the next section, in the special case of $N = 2$, $n$ is a homomorphism. Indeed, in that case it is possible to define a mod-2 reduction map
\be O(2, \mathbb{Q}) \overset{\mathbb{F}_2}{\longrightarrow} O(2, \mathbb{F}_2) \cong \mathbb{Z}_2 \nn\ee
which, when multiplied by
\be O(2, \mathbb{Q}) \overset{\det}{\longrightarrow} \{\pm 1\} \cong \mathbb{Z}_2 \nn\ee
gives a homomorphism that coincides with our $n$. (We thank Holly Krieger for this observation.) However, a clean interpretation of $n(\mathcal{R})$ for $N > 2$ is not so obvious.

\section{Examples} \label{examplesec}

In this section, we describe a number of different examples of boundary states and the resulting ground state degeneracy.

\subsection{Simple Boundary States}

The two simplest boundary conditions are the generalisations of the vector and axial conditions described in the introduction, now imposed independently on each of the $N$ fermions. These are given by
\begin{itemize}
\item Vector: $Q = \bar{Q} = \mathds{1}$. This gives $\mathcal{R} = \mathds{1}$ and $\vol(\Lambda) = 1$.
\item Axial: $Q = \mathds{1}$ and $\bar{Q} = -\mathds{1}$. This gives $\mathcal{R}= -\mathds{1}$ and $\vol(\Lambda) = 1$.
\end{itemize}
If we impose the same boundary conditions at both ends, the generic ground state degeneracy is $G[\mathcal{R},\mathcal{R}] =1$. (As we have seen, this can be enhanced for special values of the phases.)

\para
In contrast, if we impose vector boundary conditions at one end and axial boundary conditions at the other, we have $\LambdaRRdash = \{0\}$. In this case $\mathcal{R}^T \mathcal{R}' = -\mathds{1}$ and the formula \eqn{G} gives
\be \mbox{Vector-Axial:}\ \ G[\mathcal{R},\mathcal{R}'] = 2^{N/2} \nn\ee
This is the expected answer since, as explained in the introduction, this system has $N$ Majorana zero modes. This means that the vector and axial boundary conditions sit in the same class for $N$ even, but different classes for $N$ odd.

\para
There is a third interesting boundary condition which arises in the study of fermions scattering off monopoles \cite{callan,joe}. Following \cite{sagi}, we refer to it as the \emph{dyon boundary condition}. It is given by
\begin{itemize}
\item Dyon: It is simplest to specify the boundary condition in terms of the orthogonal matrix $\mathcal{R} = \bar{Q}^{-1} Q$, which is given by
\be \mathcal{R}_{ij} = \delta_{ij} - \frac{2}{N} \nn\ee
The charge lattice has $\vol(\Lambda) = N/2$ for $N$ even and $\vol(\Lambda) = N$ for $N$ odd. The corresponding charge matrices contain only $\pm 1$. For $N = 4$ they are given by
\be Q_{\alpha i} = \scaleobj{.8}{\left(\begin{array}{cccc}
+ & + & + & +\\
+ & -\\
 & + & -\\
 & & + & -
\end{array}\right)}
\qquad \text{and} \qquad
\bar{Q}_{\alpha i} = \scaleobj{.8}{\left(\begin{array}{cccc}
- & - & - & -\\
+ & -\\
 & + & -\\
 & & + & -
\end{array}\right)}
\nn\ee
with the obvious extension to general $N$.
\end{itemize}
We now have two further pairings to consider:

\para
The case of vector-dyon boundary conditions was considered in \cite{sagi}. Here the matrix $\mathcal{R}^T\mathcal{R}'$ acts as a refletion along $\frac{1}{\sqrt{N}}(1\dots1)$, which means that we have $n_- = 1$. The intersection of the charge lattices has $\vol(\LambdaRRdash) = \sqrt{N}$. The degeneracy of ground states is then
\be
\mbox{Vector-Dyon}:\ \ G[\mathcal{R},\mathcal{R}'] =
\begin{cases}
1 & \text{$N$ even} \\
\sqrt{2} & \text{$N$ odd}
\end{cases}
\nn\ee
For axial-dyon boundary conditions, we again have $\vol(\LambdaRRdash) = \sqrt{N}$. Now, however, $\mathcal{R}^T \mathcal{R}'$ differs by a minus sign from the vector-dyon case which means that $n_-=N-1$. This time, the ground state degeneracy is always an integer
\be \mbox{Axial-Dyon}:\ \ G[\mathcal{R},\mathcal{R}'] = 2^{\lceil N/2 \rceil - 1} \nn\ee
We learn that the axial and dyon boundary condition lie in the same class.

\subsection{Two Dirac Fermions} \label{twosec}

We now turn to the case of $N=2$ fermions, where we can simply enumerate all possible boundary conditions and determine their class. This extends and completes the proof in Section~\ref{twoclasses}, but only for this special case.

\para
These boundary conditions include the example given in the introduction, where left-movers have charges $(3,4)$ and right-movers have charges $(5,0)$. However, our boundary state formalism require that a $U(1)^2$ symmetry is imposed on the boundary, which means that we must supplement the charges above with a second $U(1)$ symmetry. It is straightforward to find such symmetries: for example, we can take the left-movers to have charges $(-4,3)$ and right-movers have charges $(0,5)$. Alternatively, we could take the left-movers to have charges $(4,-3)$ and the right-movers to have charges $(0,5)$. In what follows, we will see that all such boundary conditions can be associated to pythagorean triples in this way. However, rather surprisingly, the choice of the minus signs in the second $U(1)$ can dramatically change the resulting physics.

\para
We specify the boundary condition using the rational orthogonal matrix $\mathcal{R}$ defined in \eqn{R}. Such matrices are either rotations or reflections and can be written accordingly as
\be
\mathcal{R}_\text{rot} = \frac{1}{c} \left(\begin{array}{cc}
a & b \\
-b & a
\end{array}\right)
\qquad \text{or} \quad
\mathcal{R}_\text{ref} = \frac{1}{c} \left(\begin{array}{cc}
a & b\\
b & -a
\end{array}\right)
\label{rpythag}\ee
where $a,b,c$ are co-prime integers with $a^2 + b^2 = c^2$ and $c > 0$.

\para
It will be useful to first compute $\vol(\Lambda)$ for such boundary conditions. We have

\vskip 4mm
\noindent {\bf Claim:} $\vol(\Lambda) = c$

\vskip 4mm
\noindent {\bf Proof:}
Consider rotation matrices. The charge lattice $\Lambda$ consists of all integer-valued vectors $\left(\begin{smallmatrix}x \\ y \end{smallmatrix}\right)$ such that $\mathcal{R} \left(\begin{smallmatrix}x \\ y \end{smallmatrix}\right)$ is also integer-valued. In other words, we're looking for all integer solutions to
\be ax+by = cz \ \ \text{and} \ \ -bx+ay = cw \nn\ee
Since $a,b$ are coprime, there exist integers $\lambda, \mu$ such that
\be a\lambda+b\mu=1 \nn\ee
Therefore any value of $z$ can be attained by some $(x,y)$, and for fixed $z$, the possible values of $(x,y)$ are
\be (x,y) = cz(\lambda,\mu) + n(-b,a) \nn\ee
where $n$ is a free integer. Plugging this into the second equation, we then find that $w$ is automatically also an integer,
\be w = z(-b\lambda + a\mu) + cn \nn\ee
The lattice $\Lambda$ is therefore spanned by $c(\lambda,\mu)$ and $(-b,a)$. The volume of the unit cell is
\be \vol(\Lambda) = \det \! \left(\begin{array}{cc}c\lambda & -b \\ c\mu & a\end{array}\right) = c(\lambda a + \mu b) = c \label{volume}\ee
The proof for reflection matrices proceeds in an identical fashion. \hfill $\Box$

\para
Our next goal is to compute the ground state degeneracy \eqn{G} for an interval sandwiched between two boundaries $\mathcal{R}$ and $\mathcal{R}'$. As always, when $\mathcal{R} = \mathcal{R}'$, the ground state degeneracy is $G[\mathcal{R},\mathcal{R}] = 1$. The remaining cases are less trivial.

\para
First, it will prove useful to parameterise the Pythagorean triple $(a,b,c)$ in \eqn{rpythag} using Euclid's formula,
\be \mathcal{R}_\text{rot}(p,q) = \frac{1}{p^2 + q^2} \left(\begin{array}{cc}p^2 - q^2 & 2pq \\ -2pq & p^2 - q^2\end{array}\right) \label{rot}\ee
and
\be \mathcal{R}_\text{ref}(p,q) = \left(\begin{array}{cc}1 & 0 \\ 0 & -1\end{array}\right) \mathcal{R}_\text{rot}(p,q) \label{ref}\ee
with $p,q$ co-prime.

\para
Usually in applications of Euclid's formula, one further assumes that $p$ and $q$ are not both odd, which gives rise to a primitive Pythagorean triple. We do not insist on this condition here since it allows us to construct rotation matrices \eqn{rpythag} with $b$ odd. For example,
\be
p=2,\ q=1\ \ &\Rightarrow \ \ \mathcal{R}_\text{rot}(p,q) = \frac{1}{5}\left(\begin{array}{cc} 3 & 4 \\ -4 & 3 \end{array}\right) \nn\\
p=3,\ q=1\ \ &\Rightarrow \ \ \mathcal{R}_\text{rot}(p,q) = \frac{1}{5}\left(\begin{array}{cc} 4 & 3 \\ -3 & 4 \end{array}\right) \nn
\ee
Nonetheless, as we go on, we will see that the distinction between $p$ and $q$ both odd, or one odd and one even, becomes more prominent. For example, the volume of the unit cell \eqn{volume} is
\be \vol(\Lambda) = \begin{cases}p^2 + q^2 & \text{if either $p$ or $q$ is even} \\ \frac{1}{2} (p^2 + q^2) & \text{if $p$ and $q$ are both odd} \end{cases} \label{vol2}\ee
Indeed, ultimately we will see that the boundary conditions sit in one of two classes as follows:
\be \text{{Class $\classv$:}} \quad \begin{cases} \text{Rotation matrices $\mathcal{R}_\text{rot}(p,q)$ with either $p$ or $q$ even.} \\[1.2ex] \text{Reflection matrices $\mathcal{R}_\text{ref}(p,q)$ with $p$ and $q$ both odd.} \end{cases} \nn\ee
and
\be \text{{Class $\classa$:}} \quad \begin{cases} \text{Rotation matrices $\mathcal{R}_\text{rot}(p,q)$ with $p$ and $q$ both odd.} \\[1.2ex] \text{Reflection matrices $\mathcal{R}_\text{ref}(p,q)$ with either $p$ or $q$ even.} \end{cases} \nn\ee
To see this, we first consider two separate cases.
\begin{itemize}
\item \underline{Case 1:} $\det(\mathcal{R}^T \mathcal{R}') = +1$ with $\mathcal{R} \neq \mathcal{R}'$.

Here $\mathcal{R}$ and $\mathcal{R}'$ describe two different rotations or two different reflections. Either way, $\mathcal{R}^T\mathcal{R}'$ has no $+1$ eigenvalue and so $\LambdaRRdash = \{0\}$. We can then use the simplified expression \eqn{Gsimple} for the ground state degeneracy. A direct evaluation, using the form of the matrices \eqn{rpythag} gives
\be G[\mathcal{R},\mathcal{R}'] = \sqrt{2(cc'-aa'-bb')} \nn\ee
It is not at immediately obvious that this is an integer or $\sqrt{2}$ times an integer. However, invoking the Euclid form of the matrix \eqn{rot} or \eqn{ref}, it is not hard to show that the ground state degeneracy can be written as
\be G[\mathcal{R},\mathcal{R}'] = \begin{cases} 2|pq'-qp'| &\text{if $(p,q)$ and $(p',q')$ lie in the same class} \\ \sqrt{2}\,|pq'-qp'| &\text{if $(p,q)$ and $(p',q')$ lie in different classes} \end{cases} \label{2g}\ee
This confirms our classification if both matrices are rotations or both are reflections.
\end{itemize}

\para
To illustrate this, consider the following three rotation matrices
\be
\mathcal{R}_1 = \left(\begin{array}{cc}1 & 0 \\ 0 & 1\end{array}\right) \quad , \quad
\mathcal{R}_2 = \left(\begin{array}{cc}-1 & 0 \\ 0 & -1\end{array}\right)\quad , \quad
\mathcal{R}_3 = \frac{1}{5} \left(\begin{array}{cc}3 & 4 \\ -4 & 3\end{array}\right) \quad , \quad
\mathcal{R}_4 = \frac{1}{5} \left(\begin{array}{cc}4 & 3 \\ -3 & 4\end{array}\right)
\nn\ee
From the discussion above, $\mathcal{R}_1$, $\mathcal{R}_2$ and $\mathcal{R}_3$ all lie in class $\classv$ while $\mathcal{R}_4$ lies in class $\classa$. The number of ground states in an interval with one of these boundary conditions imposed on each end is
\begin{center}
  \begin{tabular}{c|cccc}
     & $\mathcal{R}_1$ & $\mathcal{R}_2$ & $\mathcal{R}_3$ & $\mathcal{R}_4$ \\
    \hline
    $\mathcal{R}_1$ & 1 & 2 & 2 & $\sqrt{2}$ \\
    $\mathcal{R}_2$ & 2 & 1 & 4 & $3\sqrt{2}$ \\
    $\mathcal{R}_3$ & 2 & 4 &1 & $\sqrt{2}$ \\
    $\mathcal{R}_4$ & $\sqrt{2}$ & $3\sqrt{2}$ & $\sqrt{2}$ & 1
  \end{tabular}
\end{center}
\vspace{1em}
\noindent
Although the number of ground states in class $\classv$ in these examples have the form $2^n$, as is familiar from quantising complex fermionic zero modes, it is clear from the general form \eqn{2g} that we can get any even number of ground states in this class of examples.

\para
The second case corresponds to one of the special cases not handled by the proof in Section~\ref{twoclasses}, and requires a little more work. This is
\begin{itemize}
\item \underline{Case 2:} $\det(\mathcal{R}^T \mathcal{R}') = -1$

Here one of $\mathcal{R}$ and $\mathcal{R}'$ describes a rotation and the other a reflection. This means that the eigenvalues of $\mathcal{R}^T \mathcal{R}'$ are $+1$ and $-1$, and so $\detprime(\mathds{1} - \mathcal{R}^T \mathcal{R}') = +2$.

The single $+1$ eigenvalue of $\mathcal{R}^T \mathcal{R}'$ implies that $\LambdaRRdash$ is one-dimensional. We must compute the volume of the unit cell of this lattice and this is a little involved. Without loss of generality, we take $\mathcal{R}_\text{rot}[p,q]$ and $\mathcal{R}'_\text{ref}(p',q')$. The unique $+1$ eigenvector of $\mathcal{R}^T \mathcal{R}'$ is, up to proportionality,
\be
v = \left(\begin{array}{c}pp'-qq' \\ pq'+qp'\end{array}\right)
\ \ \ \Rightarrow\ \ \
\mathcal{R}_\text{rot}(p,q) v = \mathcal{R}_\text{ref}(p',q')v = \left(\begin{array}{cc}pp' + qq' \\ pq'-qp'\end{array}\right)
\nn\ee
Clearly, both $v$ and $\mathcal{R}_\text{rot} v = \mathcal{R}'_\text{ref} v$ are integer vectors. The trouble lies in the caveat of proportionality: it may be possible to divide $v$ by some integer $d$ so that the conclusion that we have an integer-valued eigenvector remains true. In fact, such a $d$ is simply the greatest common divisor of the four components of $v$ and $\mathcal{R}_\text{rot}v$,
\be d = \gcd(pp'-qq' \,;\, pq'+qp' \,;\, pp'+qq' \,;\, pq'-qp') \nn\ee
We have $d=2$ if $p,q,p',q'$ are all odd; otherwise $d=1$. The one-dimensional lattice $\LambdaRRdash$ is then spanned by the single vector $v/d$, and we have
\be
\vol(\LambdaRRdash) = \frac{|v|}{d} = \frac{\sqrt{(p^2 + q^2)(p^{\prime 2} + q^{\prime 2})}}{d}
\nn\ee
We now have all the information we need to compute the ground state degeneracy \eqn{G}. Using the expression \eqn{vol2} for the volume of the unit cells, we have
\be G[\mathcal{R}_\text{rot}, \mathcal{R}'_\text{ref}] = \begin{cases}1 & \text{if $\mathcal{R}_\text{rot}$ and $\mathcal{R}'_\text{ref}$ belong to the same class} \\ \sqrt{2} & \text{if $\mathcal{R}_\text{rot}$ and $\mathcal{R}'_\text{ref}$ belong to different classes}\end{cases} \nn\ee
\end{itemize}

\newpage
\appendix

\section{Fermion Conventions} \label{fermionsappendix}

Our convention for a Majorana fermion in 1+1D is
\be S = \frac{i}{4\pi} \int \! dx dt \left( \chi_+ \partial_+ \chi_+ + \chi_- \partial_- \chi_- \right) \nn\ee
where $\partial_\pm = \partial_t \pm \partial_x$. This Euclideanises to the standard CFT action
\be S = \frac{1}{2\pi} \int \! dx d\tau \left( \chi \bar{\partial} \chi + \bar{\chi} \partial \bar{\chi} \right) \nn\ee
where $z = x + i \tau$, provided we define the new fermions by
\be \chi = e^{-i\pi/4} \, \chi_+ \ \ \ \text{and} \ \ \ \bar{\chi} = e^{+i\pi/4} \, \chi_- \nn\ee
The $N$ Dirac fermions are built from $2N$ Majorana fermions via $\psi_i = \frac{1}{\sqrt{2}}(\chi_{2i-1} + i \chi_{2i})$. The corresponding $U(1)$ currents are
\be J_i = i \chi_{2i-1} \chi_{2i} \nn\ee

\section{Clustering and the Cardy-Lewellen Sewing Conditions} \label{sewingsec}

In the rest of this appendix we describe a few subtleties that we felt were best avoided in the main text.

\para
In Section~\ref{ishibashisec}, a set of ground states $\ket{\lambda, \bar{\lambda}}$ was introduced, but at the time we did not specify their phases. The easiest way to do this is via the bosonisation formula,\footnote{Our handling of Klein factors takes inspiration though slightly differs from \cite{delft}.}
\be \psi_i(z) = F_i \, t_i \, z^{-\lambda_i} \exp\!\left( -\sum_{n=1}^\infty \frac{z^n}{n} J_{i,-n} \right) \exp\!\left( \sum_{n=1}^\infty \frac{z^{-n}}n J_{i,n} \right) \nn\ee
Here, $F_i$ is a ladder operator which moves between ground states as $F_i \ket{\lambda,\bar{\lambda}} = \ket{\lambda - \hat{e}_i, \bar{\lambda}}$, and $t_i$ is a cocycle arising from Fermi statistics which acts by a phase on each ground state. The precise form of $t_i$ (and its barred cousin $\bar{t}_i$) will depend on the phase convention chosen for the $\ket{\lambda,\bar{\lambda}}$. We stipulate them to be
\be t_i = (-1)^{\sum_{j=1}^{i-1} \lambda_j} \ \ \ \text{and} \ \ \ \bar{t}_i = (-1)^{\sum_{j=1}^{i-1} \bar{\lambda}_j + \sum_{j=1}^N \lambda_j} \nn \ee
and this then implicitly fixes the relative phases of the $\ket{\lambda,\bar{\lambda}}$.

\para
In Section~\ref{cardysec}, it was claimed that the requirement of cluster decomposition in the presence of the boundary state $\ket{a; \mathcal{R}}$ dictates the form of the coefficients $a_\lambda$. Here we give more details.

\para
To formulate the requirement of clustering, we start by placing the theory on the planar region $|z| \geq 1$, and impose the boundary condition $\ket{a; \mathcal{R}}$ at $|z| = 1$. Let $\mathcal{O}_i(z)$ be a list of all composite local operators built out of the fermions.\footnote{To lighten the notation, we have restricted to the real slice $\bar{z} = z^*$, so $\mathcal{O}_i(z)$ should not be interpreted as a purely holomorphic operator.} Then we demand that the following limit involving a ratio of normalised correlators is equal to one,
\be \lim_{|z|, |w| \rightarrow 1^+} \frac{\langle \mathcal{O}_i(z) \mathcal{O}_j(w) \rangle}{\langle \mathcal{O}_i(z) \rangle \langle \mathcal{O}_j(w) \rangle} = 1 \nn\ee
where the limit is taken with $\arg(z)$ and $\arg(w)$ fixed.

\para
The $\mathcal{O}_i(z)$ must have non-vanishing vev in the presence of the boundary. This condition will be met if our operator has compatible $U(1)^N$ charges $(q, \bar{q})$, in the sense that $\bar{q} = -\mathcal{R} q$. In particular, we are forced to take $q \in \LambdaR$. To build an operator with these charges, we can take $|q_i|$ copies of each $\psi_i(z)$, and $|\bar{q}_i|$ copies of each $\bar{\psi}_i(z)$, and combine them into a composite operator $\mathcal{O}_q(z)$ using a suitable point-splitting regularisation. (If any of the charges $q_i$ are negative, we should replace $\psi_i$ with its complex conjugate, $\frac{1}{\sqrt{2}}(\chi_{2i-1} - i \chi_{2i})$.) The clustering requirement for $\mathcal{O}_q$ and $\mathcal{O}_p$ is then
\be \lim_{|z|, |w| \rightarrow 1^+} \frac{\bra{0, 0} \mathcal{O}_q(z) \mathcal{O}_p(w) \ket{a; \mathcal{R}} \; \braket{0, 0}{a; \mathcal{R}}}{\bra{0, 0} \mathcal{O}_q(z) \ket{a; \mathcal{R}} \; \bra{0, 0} \mathcal{O}_p(w) \ket{a; \mathcal{R}}} = 1 \nn\ee
It turns out that the only interesting contribution to this expression comes from the $F_i \, t_i$ part of $\psi_i(z)$, and everything else can dropped. That is, we can make the replacement
\be \mathcal{O}_q(z) \; \longrightarrow \; \prod_{i=1}^N (F_i t_i)^{q_i} \, \prod_{i=1}^N (\bar{F}_i \bar{t}_i)^{\bar{q}_i} \nn\ee
whereupon the clustering condition reduces to
\be \frac{a_{q+p} \, a_0}{a_q \, a_p} \frac{\bra{q, \bar{q}} \mathcal{O}_p \ket{q + p, \bar{q} + \bar{p}}}{\bra{0, 0} \mathcal{O}_p \ket{p, \bar{p}}} = 1 \nn\ee
The ratio of matrix elements in the above expression evaluates to $(-1)^{f_{\mathcal{R}}(q,p)}$ where
\be f_{\mathcal{R}}(q, p) \coloneqq \sum_{i=1}^N p_i \sum_{j=1}^{i-1} q_j + \sum_{i=1}^N (\mathcal{R} p)_i \bigg(\sum_{j=1}^{i-1}(\mathcal{R} q)_j + \sum_{j=1}^N q_j \bigg) \mod 2 \nn\ee
This is a symmetric bilinear form on $\LambdaR$ taking values mod 2, whose corresponding quadratic form is fermion parity: $f_{\mathcal{R}}(\lambda, \lambda) = \lambda^2 \mod 2$. The clustering condition now takes the final form
\be \frac{a_{q+p}}{a_0} = \frac{a_q}{a_0} \frac{a_p}{a_0} (-1)^{f_{\mathcal{R}}(q,p)} \label{clusterfinal}\ee
To solve it, let $\hat{f}_\mathcal{R}(q, p)$ be an arbitrary choice of lift of $f_\mathcal{R}(q, p)$ from a mod-2 to a mod-4 valued symmetric bilinear form. Then the general solution to \eqn{clusterfinal} is
\be \frac{a_\lambda}{a_0} = e^{i \gamma_\mathcal{R}(\lambda)} e^{i \theta \cdot \lambda} \quad \text{where} \quad e^{i \gamma_\mathcal{R}(\lambda)} \coloneqq i^{\hat{f}_\mathcal{R}(\lambda, \lambda)} \nn\ee
Due to the freedom of choice in the lift $\hat{f}_\mathcal{R}(q,p)$, the reference solution $e^{i \gamma_\mathcal{R}(\lambda)}$ is actually ambiguous up to multiplication by $(-1)^{s \cdot \lambda}$ for any $s \in \LambdaR^\star$. The ambiguity is equivalent to that of choosing a quadratic refinement of $(-1)^{f_\mathcal{R}(q,p)}$, and there is no canonical way to fix it. As a result, the origin of $\theta$ is also ambiguous up to shifts by $\pi \, \LambdaR^\star$. On the other hand, the square of the reference solution is well-defined, and is equal to $(e^{i \gamma_\mathcal{R}(\lambda)})^2 = (-1)^{f_\mathcal{R}(\lambda, \lambda)} = (-1)^{\lambda^2}$.

\para
Finally, in Section~\ref{gsec}, we needed the fact that if $\mathcal{R}$ and $\mathcal{R}'$ are two different symmetries, then for all $\lambda \in \LambdaRRdash$,
\be \frac{e^{i\gamma_\mathcal{R}(\lambda)}}{e^{i\gamma_{\mathcal{R}'}(\lambda)}} = (-1)^{s \cdot \lambda} \nn\ee
for some $s \in \LambdaRRdash^\star$. (The precise value of $s$ is actually ambiguous, for the reasons described above.\footnote{One might hope that the ambiguities in $\gamma_\mathcal{R}$ could be chosen in such a way that $s$ is always zero, but sadly this turns out not to be possible.}) To see this, first note that from \eqn{clusterfinal},
\be \frac{e^{i\gamma_\mathcal{R}(q+p)}}{e^{i\gamma_{\mathcal{R}'}(q+p)}} = \frac{e^{i\gamma_\mathcal{R}(q)}}{e^{i\gamma_{\mathcal{R}'}(q)}} \frac{e^{i\gamma_\mathcal{R}(p)}}{e^{i\gamma_{\mathcal{R}'}(p)}} \frac{(-1)^{f_{\mathcal{R}}(q,p)}}{(-1)^{f_{\mathcal{R}'}(q,p)}} \nn\ee
and that $f_{\mathcal{R}}(q,p) = f_{\mathcal{R}'}(q,p)$ for $q,p \in \LambdaRRdash$. This forces $\frac{e^{i\gamma_\mathcal{R}(\lambda)}}{e^{i\gamma_{\mathcal{R}'}(\lambda)}}$ to take the form $e^{i \theta \cdot \lambda}$. Since it also squares to $\frac{(-1)^{\lambda^2}}{(-1)^{\lambda^2}} = 1$, we must have $\theta \in \pi \, \LambdaRRdash^\star$.

\section{Lattice Calculations} \label{latticesappendix}

We record here a technical calculation of lattice volumes that we used several times in Section~\ref{twoclasses}. Let $N$ be an even number, let $A$ be a rational $N \times N$ antisymmetric matrix, and let $\mathcal{R} = \frac{\mathds{1} - A}{\mathds{1} + A}$. Then we claim that
\be \vol(\LambdaR) = 2^{-n} \det(\mathds{1} + A) \, \prod_{i=1}^{N/2} (g/g_{i})^2 \nn\ee
where the integers $n, g, g_i$ are constructed along the way during the proof.

\vskip 4mm
\noindent {\bf Proof:}
To describe $\LambdaR$, we need to find all integer solutions to
\be \mathcal{R} v = w \nn\ee
In terms of new variables $x = v - w$ and $y = v + w$, this reads
\be x = Ay \nn\ee
Let us pull out a common denominator from $A$ by writing it as $A = \tilde{A}/g$ with $\tilde{A}$ an integer matrix. We also invoke the Smith-like decomposition of $\tilde{A}$,
\be \tilde{A} = U \! D U^T \qquad D = J \, \text{ddiag}(\nu_i) \qquad J = \bigoplus_{i=1}^{N/2} \left(\begin{array}{cc}0 & 1\\-1 & 0\end{array}\right) \nn\ee
with $U$ unimodular and $\nu_i$ integers. Then in terms of further new variables $\tilde{x} = U^{-1} x$ and $\tilde{y} = U^T y$, which are still integer vectors, our equation becomes
\be g \tilde{x} = D \tilde{y} \nn\ee
which is now diagonal, hence trivial to solve. The set of all solutions can be parametrised, in terms of an integer vector $z$, via
\be \tilde{x} = J \, \text{ddiag}(\nu_i/g_i) z \qquad \tilde{y} = \text{ddiag}(g/g_i) z \nn\ee
with $g_i = \gcd(g, \nu_i)$. Returning to the original variable $v$, we have
\be 2v = Xz \qquad X = U^{T,-1} \text{ddiag}(g/g_{i}) + U J \, \text{ddiag} (d_i/g_i) \nn\ee
We are almost done, except for the requirement that $v$ be integral, which places a constraint on the allowed values of $z$:
\be Xz = 0 \mod 2 \nn\ee
By considering the SNF of $X$, one can show that this constraint forces $z$ to lie in a certain sublattice $\Lambda_z \subseteq \mathbb{Z}^N$, whose volume is
\be \vol(\Lambda_z) = 2^{N-n} \nn\ee
where $n = \text{nullity}_{\,\mathbb{F}_2}(X)$. With $z$ and $v$ now integer vectors, $w$ automatically is too, and so we have parametrised all integer solutions to $\mathcal{R} v = w$. The lattice $\LambdaR$ is then the set of allowed values of $v$. To calculate its volume, we simply chain together several earlier equations, namely $v = \tfrac{1}{2} (\mathds{1} + A) y$, $y = U^T \tilde{y}$, $\tilde{y} = \text{ddiag}(g/g_i) z$, and $z \in \Lambda_z$, with the result
\be
\vol(\LambdaR) =
\det \!\big( \tfrac{1}{2} (\mathds{1} + A) \big) \,
\det \!\big( U^T \big) \,
\det \!\big( \text{ddiag}(g/g_i) \big) \,
\vol   \big( \Lambda_z \big)
\nn\ee
which gives the formula stated at the beginning. An entirely analogous result also holds for odd $N$, with an identical proof.

\section{Boundary States for a Majorana Fermion}

The theory of boundary conditions for Virasoro minimal models is well-understood, where it is known that the overlap between any two boundary states gives rise to an integer ground state degeneracy \cite{cardy}. However, this appears to be at odds with the situation for a Majorana fermion, which is also described by a minimal model $\mathcal{M}(4, 3)$, yet depending on the boundary conditions, may give rise to a ground state degeneracy of $\sqrt{2}$. The loophole is that the first statement only holds for modular-invariant theories, whereas a fermionic theory cannot be modular invariant, as by definition it depends on a choice of spin structure. In this section, we check that there is indeed no contradiction: treated appropriately, the $\mathcal{M}(4, 3)$ theory yields the degeneracies we expect for a Majorana fermion on an interval.

\para
As discussed in the introduction, we start by placing the Majorana fermion on a periodic annulus. The Hilbert space content is
\be \mathcal{H}_P = (\mathcal{M}_0 \oplus \mathcal{M}_{1/2}) \otimes \overline{(\mathcal{M}_0 \oplus M_{1/2})} \nn\ee
where the $\mathcal{M}_h$ are irreducible Verma modules. A clearer way to represent this information is to use a table, showing the combinations of primary fields of the underlying $\mathcal{M}(4,3)$ minimal model that occur for the Majorana fermion:
\begin{center}
\begin{tabular}{c|ccc}
${}_h \backslash {}^{\bar{h}}$ & $0$ & $\frac{1}{2}$ & $\frac{1}{16}$ \\
\hline
$0$ & $\bullet$ & $\bullet$ & \\
$\frac{1}{2}$ & $\bullet$ & $\bullet$ & \\
$\frac{1}{16}$ & & & \\
\end{tabular}
\end{center}
Ishibashi states can only come from the diagonal, so there are two of them: $\kket{0}$ and $\kket{\frac{1}{2}}$. Let us now form two putative boundary states
\be
\ket{a} &= a_0 \kket{0} + a_{1/2} \kket{\tfrac{1}{2}} \nn\\
\ket{b} &= b_0 \kket{0} + b_{1/2} \kket{\tfrac{1}{2}} \nn
\ee
with arbitrary complex coefficients. The partition function \eqn{closed} between them is
\be
\mathcal{Z}_P &= \bra{b} (-1)^F q^{\frac{1}{2}(L_0 + \bar{L}_0 - \frac{c}{12})} \ket{a} \nn\\
&= \bar{b}_0 a_0 \, \chi_0 - \bar{b}_{1/2} a_{1/2} \, \chi_{1/2} \nn
\ee
The factor of $(-1)^F$ was discussed in Section~\ref{cardysec}. Here it flips the sign of the Ishibashi state $\kket{\tfrac{1}{2}}$. The modular $\mathcal{S}$-matrix for $\mathcal{M}(4,3)$ can be found, for example, in \cite{yellow}: it is
\be
S = \left(\begin{array}{ccc}
\frac{1}{2} & \frac{1}{2} & \frac{1}{\sqrt{2}} \\
\frac{1}{2} & \frac{1}{2} & -\frac{1}{\sqrt{2}} \\
\frac{1}{\sqrt{2}} & -\frac{1}{\sqrt{2}} & 0
\end{array}\right)
\nn\ee
We use this to $\mathcal{S}$-transform the previous expression $\mathcal{Z}_P$. The result is the interval partition function corresponding to boundary conditions $a$ and $b$,
\be \mathcal{Z}_{AB} = \frac{\bar{b}_0 a_0 - \bar{b}_{1/2} a_{1/2}}{2} \big( \chi_0 + \chi_{1/2} \big) + \frac{\bar{b}_0 a_0 + \bar{b}_{1/2} a_{1/2}}{2} \big( \!\sqrt{2}\, \chi_{1/16} \big) \nn\ee
If we guess the fundamental boundary states to have the form
\be \ket{\pm} = \kket{0} \pm \kket{\tfrac{1}{2}} \nn\ee
then their interval partition functions are
\be
\mathcal{Z}_{++} = \mathcal{Z}_{--} &= \sqrt{2} \, \chi_{1/16} \nn\\
\mathcal{Z}_{+-} = \mathcal{Z}_{-+} &= \chi_0 + \chi_{1/2} \nn
\ee
The interpretation is that the boundary states $\ket{\pm}$ simply correspond to the two possible boundary conditions $\psi_L = \pm \psi_R$ one can impose on a Majorana fermion. To see this is the correct interpretation, we need the identities
\be
\chi_{1/16} &= q^{1/24} \prod_{n=1}^\infty (1 + q^n) \nn\\
\chi_0 + \chi_{1/2} &= q^{-1/48} \prod_{n=1/2}^\infty (1 + q^n) \nn
\ee
These expressions are very clearly the partition functions of the non-zero modes of a Majorana fermion with boundary conditions $++$ and $+-$ respectively. For the $++$ case, $\mathcal{Z}_{++}$ also contains an extra factor of $\sqrt{2}$ on top of $\chi_{1/16}$, which we must interpret as the contribution from the single zero mode.

\para
The boundary state formalism appears to have singled out the convention that an unpaired Majorana mode contributes $\sqrt{2}$ to the partition function. The reason is that this is the only way for the theory to give the right answer for an even number of copies of the system, as the partition function must simply scale extensively with the number of fermions. We conclude that the normalisations of the boundary states $\ket{\pm}$ are appropriate for describing the theory of a Majorana fermion, and, at least within this context, it's acceptable for Cardy's condition to involve factors of $\sqrt{2}$ rather than integers, which is only possibile at all due to the modular non-invariance of the theory.

\acknowledgments

We thank Nick Dorey, Kristan Jensen, Anton Kapustin, Charles Kane, Curt von Keyserlingk, Holly Krieger, Shinsei Ryu, John Terning, Juven Wang and Gerard Watts for useful discussions. We are supported by the STFC consolidated grant ST/P000681/1. DT is a Wolfson Royal Society Research Merit Award holder and is supported by a Simons Investigator Award. PBS would like to thank the Cambridge Trust for support through the Vice-Chancellor's Award.

\end{document}